\documentclass[12pt,a4paper,english,superscriptaddress,aps,nofootinbib]{revtex4}
\usepackage[utf8]{inputenc}
\usepackage[T1]{fontenc}
\usepackage{amsmath,amssymb,graphicx}
\makeatletter
\usepackage{babel}
\usepackage[active]{srcltx}
\usepackage{graphicx,color}
\usepackage{changebar}
\usepackage{hyperref}
\usepackage[T1]{fontenc}
\usepackage{esint}
\usepackage{multirow}
\usepackage{dsfont}
\usepackage{ae}
\usepackage{amsmath}
\usepackage{braket}
\usepackage{mathtools}
\usepackage{slashed}
\usepackage{empheq}
\usepackage{multirow}
\usepackage{graphicx}
\usepackage{amsfonts}
\usepackage{amsmath}
\usepackage{caption}
\begin{document}
\title{Weak gravitational lensing by an ESTGB  black hole in the presence of a plasma}

\author{Qian Li}
\affiliation{Faculty of Science, Kunming University of Science and Technology, Kunming, Yunnan 650500, China.}

\author{Yu Zhang}
\email{zhangyu\_128@126.com  (Corresponding author)}
\affiliation{Faculty of Science, Kunming University of Science and Technology, Kunming, Yunnan 650500, China.}

\author{Zhi-Wen Lin}
\affiliation{Faculty of Science, Kunming University of Science and Technology, Kunming, Yunnan 650500, China.}

\author{Qi-Quan Li}
\affiliation{Faculty of Science, Kunming University of Science and Technology, Kunming, Yunnan 650500, China.}

\author{Qi Sun}
\affiliation{Faculty of Science, Kunming University of Science and Technology, Kunming, Yunnan 650500, China.}

%%%%%%%%%%%%%%%%%%%%%%%%%%%%%%%%%%%%%

%%%%%%%%%%%%%%%%%%%%%%%%%%%%%%%%%%%%%

\begin{abstract}
		This paper is devoted to studying the weak-field gravitational lensing properties of a 4D ESTGB  black hole, which is surrounded by the plasma medium. The effects of the magnetic charges and the three plasma distribution models in the deflection of light around a 4D  ESTGB  black hole are investigated in detail.  We find that the uniform plasma leads to a larger deflection of light rays in comparison with the singular isothermal sphere (SIS), the non-singular isothermal sphere (NSIS) models. Moreover, the deflection angle increases slightly as the absolute value of the magnetic charge decreases. Finally, we analyze the total magnification of image due to weak gravitational lensing around the black hole. The result shows that the presence of a uniform plasma medium remarkably enhances the total magnification whereas the non-uniform plasma reduces the total magnification.
\end{abstract}

\maketitle

\textit{Keywords}: Black hole, Weak graviatational lensing, Plasma

PACS numbers: 04.70.Dy, 04.50.Kd, 03.65.Xp

\section{Introduction}
As one of Einstein's general relativity  predictions, black holes are the most mysterious objects in the present universe. Because the light ray is unable to escape the event horizon, which is a one-way causal boundary, black holes are not visible objects, and their existence can only be proven indirectly. However, with the development of related astronomical technology, the EHT cooperation organization \cite{Akiyama2019} published the shadow of a supermassive black hole in 2019. This may be another powerful evidence of the existence of black holes after LIGO-Vigro detected the gravitational wave signals generated by the merger of  binary black holes \cite{LIGOScientific:2016aoc}. In addition to  the standard general relativity,  many modified gravity theories are proposed  due to fundamental general relativity may not hold in high- or low- curvature regimes, such as the extended scalar-tensor-Gauss-Bonnet (ESTGB) theory  \cite{Doneva:2018rou}.  It is given through the coupling of the Gauss-Bonnet invariant with a scalar field owing to avoidance of  Ostrogradski instability, which is a special and interesting extension.  This modified theory is a natural modification of  general relativity and  extension of the standard scalar-tensor theory. The Doneva and Yazadjiev indicated that below a certain critical mass, the Schwarzschild spacetime becomes unstable in ESTGB gravity \cite{Doneva:2017bvd}.  The ESTGB theory can explain the phenomenon of the present stage of cosmic acceleration in cosmology \cite{Heydari-Fard:2016nlj}. Shortly thereafter, Ca\~nate and Perez Bergliaffa  \cite{Canate:2020kla} proposed the first exact magnetic black hole solution based on the extended scalar-tensor-Gauss-Bonnet theory (ESTGB) with a special type of nonlinear electrodynamics. The ESTGB black hole solution is characterized by the Arnowitt-Deser-Misner (ADM) mass and magnetic charge. When $m>0$ and $q<0$, the black hole solution is similar to the  Reissner-Nordstr$\rm\ddot{o}$m black hole solution.	The gray-body factor and absorption cross section of the massless Dirac field for this black hole were studied in Ref.\cite{Li:2022jda}. Ma et al. \cite{Ma:2022gzr} investigated the quasinormal modes and absorption cross section of the massless scalar field for this black hole. Besides, the thermodynamical properties for this black hole under the generalized uncertainty principle (GUP) have been studied in Ref.\cite{Lin:2022eix}.

Because the spacetime around compact massive objects is curved, one of the remarkable characteristics of general relativity is light deflection and the lens effect.  The phenomenon of light deflection and lens effect is called gravitational lensing. One of the three well-known verification experiments for general relativity involves light deflection. Therefore, gravitational lensing is used as a special tool to verify whether the general relativity theories are correct and to probe  properties of matter surrounding black hole. Besides, one can obtain some feature information of the gravitational object by the gravitational lensing. It is extremely important that the difference between different black hole lenses can be obtained by the gravitational lensing effect \cite{Eiroa:2005ag,Wei:2011bm}. So gravitational lensing still is the very active research area in the weak and strong field limits. The weak deflection angle of Schwarzschild spacetime in vacuum can be expressed by in form $\hat\alpha=2R_{s}/b$ where $R_{s} =2M $ and $b$ is the impact parameter.  Virbhadra et al. studied the  strong gravitational lensing in the context of  Schwarzschild black hole \cite{Virbhadra:1999nm}. The variation of the tangential, radial, and total magnification of the images with respect to the angular source position is investigated by simulating the supermassive black holes M87* as a Schwarzschild lens \cite{Virbhadra:2022iiy}. Sereno \cite{Sereno:2003nd} obtained the time delay and deflection angle expressions of the Reissner-Nordstr$\rm\ddot{o}$m black holes under the weak field approximation. In addition, many attempts  have been made on  the weak  deflection angle of the different modified gravity theories  by using  different methods  \cite{Jusufi:2017vta,Ovgun:2018oxk,Li:2020wvn,Fu:2021akc,Javed:2020pyz,Javed:2021arr,Li:2021xhy,Crisnejo:2019xtp,Crisnejo:2019ril,Jha:2021eww}. Generally, the angle of deflection or the relevant optical scalar can be expressed in the form of derivatives of the different components of the black hole metric. In strong gravity field, the study of gravitational lensing is a trending topic. There have been a number of articles  examining the gravitational lensing in the strong field \cite{Virbhadra:2002ju,Rahvar:2018nhx,Bozza:2010xqn,Virbhadra:2008ws,Chen:2013vja,Ji:2013xua,Chen:2015cpa,Chen:2016hil,Zhang:2017vap,Abbas:2019olp,Abbas:2021whh,Hensh:2021nsv}.

On the other hand, it is believed that compact astrophysical objects are immersed in a complicated environment, such as  plasma. In this paper, we only focus on the plasma environment. Plasma is a dispersive medium whose refractive index relies on the frequency of photons. The plasma around compact astrophysical objects affects the trajectories of the light ray since it may interact with electromagnetic waves.  Synge \cite{Synge:1960ueh} firstly proposed the self-consistent  approach to the propagation of light rays in the gravitational field in the context of plasma medium. Forty years later, Perlick \cite{Perlick2000} proposed a different type of the method  to obtain the integral expression of the deflection angle as the plasma surrounds the Schwarzschild and Kerr black holes. Later, Bisnovatyi-Kogan and Tsupko \cite{Bisnovatyi-Kogan:2008qbk} found that the deflection angle relies on the photon frequency in the uniform dispersive medium. The phenomenon has qualitatively different from the vacuum environment. The authors \cite{Bisnovatyi-Kogan:2010flt} also considered the case that the gravitational object is surrounded by the inhomogeneities of plasma and obtained the expression for the deflection angle of the different plasma models. Schee \cite{Schee:2017hof}  et al. studied the gravitational lensing about the regular black hole  immersed in plasma.  The weak deflection angle of the wormhole solution described by  exponential metric was obtained in Ref.\cite{Turimov:2022iff}. The influences of uniform plasma on the the shadow and weak deflection angle for a rotating and regular black hole in  a non-minimally coupled  Einstein-Yang-Mills (EYM) theory have been studied \cite{Kala:2022uog}.   Zhang et at.  \cite{Zhang:2022osx} studied the influences of the plasma with the power-law distribution and logarithmic normal distribution on the shadow of the Kerr black hole. In addition, Atamurotov and his coworkers were devoted to studying the weak gravitational lensing effect in plasma for various kinds of spacetimes such as  the  Lorentzian wormhole spacetime \cite{Atamurotov:2021byp}, Schwarzschild-MOG black hole \cite{Atamurotov:2021qds}, 4D Einstein-Gauss-Bonnet gravity \cite{Babar:2021exh}, rotating Einstein-Born-Infeld black hole \cite{Babar:2021nst}.

In this study, we focus on the exact expression of the deflection angle for the (3+1)-dimensional ESTGB  black hole assuming that the black hole is immersed in a plasma medium. And as an application, we will study the magnification of image in the weak field. The structure of this paper is as follows.   Section \ref{2} presents a brief  review of the process of obtaining the deflection angle under the weak-field approximation and calculating the deflection angle for the 4-dimensional  ESTGB  black hole, which is surrounded by three different plasma density distributions. In Section \ref{3}, as a type of application, we study the magnification of image for  three different plasma density distributions, i.e., uniform plasma, SIS and NSIS medium. Finally, we give our concluding remarks in Section \ref{4}.

Throughout, our choice of a spacetime  signature is $\{-,+,+,+\}$ and natural units $c = G = \hbar = 1$.  Latin indices run from 1 to 3 as well as Greek denotes from 0 to 3.

\section{Weak-field lensing in the presence of plasma}\label{2}

In this section, we will study optical properties, namely, gravitational lensing which is in the context of a 4D  ESTGB  black hole encompassed by the plasma medium under the weak-field approximation.

The 4D ESTGB gravity with an extra matter field, namely a model of non-linear electrodynamics (NLED), has the following action  \cite{Canate:2020kla}
\begin{equation}\label{0}	
S = \int d^{4}x \sqrt{-g} \bigg\{\frac{1}{4\pi} \bigg(\frac{1}{4}\big(R - \frac{1}{2}\partial_{\mu}\phi\partial^{\mu}\phi + \boldsymbol{f}(\phi) R_{_{GB}}^{2} -2 \cal{U} (\phi)\big)-\mathcal{L}_{\rm matter}\bigg) \bigg\}.
\end{equation}

Here the  first term is  the Einstein-Hilbert Lagrangian density, which is defined by the  Ricci scalar $R$, the  kinetic term of the scalar field $\frac{1}{2}\partial_{\mu}\phi\partial^{\mu}\phi$, the non-minimal coupling between the  Gauss-Bonnet invariant $R_{_{GB}}^{2}$ and scalar field $\boldsymbol{f}(\phi)$, i.e., $\boldsymbol{f}(\phi) R_{_{GB}}^{2}$, and the scalar field potential $\cal{U} (\phi)$. The Lagrangian density $\mathcal{L}_{\rm matter}$  denotes any matter field in the action. Concretely, the  Gauss-Bonnet invariant satisfies the form $R_{_{GB}}^{2}=R_{\alpha\beta\mu\nu}^{\alpha\beta\mu\nu} - 4 R_{\alpha\beta}R^{\alpha\beta}+R ^{2}$. The function $\boldsymbol{f}(\phi)$ and the scalar field potential $ \cal U (\phi)$ can be expressed as
\begin{equation}\label{00}	
\boldsymbol{f}=-\frac{\ell^{2}\sigma}{32}\!\!\left\{ \sqrt{2\sigma}\tan^{\!^{\!-1}}\!\!\!\left( \frac{\sqrt{2} }{ \sqrt{\sigma}\hskip.06cm \phi}\right) +\frac{1}{2\phi}\ln\!\!\left[\!\left(\!\!\frac{ 2\beta }{ \sigma \phi^{2} } \!+\! \beta\!\!\right)^{\!\!\!2} \right] - \frac{2}{\phi} \right\}\!\!,
\end{equation}
\begin{equation}\label{000}
\begin{aligned}
\mathcal{U}(\phi)=\frac{2^{\!^{\frac{9}{2}}}}{105\ell^{2}\sigma^{\frac{7}{2}}} \left[\frac{\pi}{2}-\!\tan^{\!^{\!-1}}\!\!\!\left(\!\!\frac{\sqrt{2} }{ \sqrt{\sigma}\hskip.06cm \phi}\!\!\right)\right]\frac{\phi^{5}}{4\ell^{2}}\left(\!\frac{3}{10\sigma}\!+\!\frac{5\phi^{2}}{7}\!+\!\frac{7\sigma\phi^{4}}{24} \!\right) \ln\!\!\left[\!\!\left(\!\!\frac{ 2\beta }{ \sigma \phi^{2} } \!+\! \beta \!\!\right)^{\!\!\!2}\right] \\
-\frac{\phi}{3\ell^{2}} \left(\!\frac{16}{35\sigma^{3}} \!-\! \frac{8\phi^{2}}{105\sigma^{2}}\!+\! \frac{31\phi^{4}}{70\sigma}\!+\! \frac{11\phi^{6}}{28}\!\right).
\end{aligned}
\end{equation}

The  NLED Lagrangian term that reduces to Maxwell's electrodynamics in the weak field regime has the following form
\begin{equation}\label{Lminder}
\mathcal{L_\text{{NLED}}}\!=\!\frac{\mathcal{F}}{8}-s^{^{\frac{1}{2}}}\!\!\!\left( 1 \!+\! \frac{37}{210\sigma_{\!\ast}} \!+\! \frac{2}{525\sigma_{\!\ast}}\!\right)\!\mathcal{F}^{^{\frac{5}{4}}} - \frac{ \sigma_{\!\ast} s \mathcal{F}^{^{\frac{3}{2}}}}{16}
+\mathcal{O}(\mathcal{F}^{^{\frac{7}{4}}}),
\end{equation}
with the  electromagnetic invariant $ \mathcal{F}=\frac{q^2}{r^4}$. And the  above parameters  have the relations $\sigma=\sigma_{*}, l=s=q, \beta=\beta_{*}$ and $\phi(r)=q/r$.

The metric describing the 4D ESTGB  black hole can be written as
\begin{equation}\label{Q1}
ds^2=-f(r)dt^2+f^{-1}(r)dr^2+r^2d\theta^2+r^2 \sin^{2}\theta d\phi^2,
\end{equation}
with
\begin{equation}\label{Q2}
f(r)=1-\frac{R_{s}}{r}-\frac{q^{3}}{r^{3}},
\end{equation}
where $R_{s}=2M$, $ M $ is ADM mass  and  $q$ is magnetic charge.

Since the weak energy condition (WEC) should be satisfied by both the corresponding effective energy-momentum tensor and that of nonlinear electrodynamics, the value of $q<0$ is permitted. Without losing generality, we consider the case that is a non-extreme black hole.  This means that the value of the magnetic charge is limited to this range $-2^{5/3}/3 < q < 0 $ when $M$ is set to 1.

We know that photons will follow the null geodesics of the effective spacetime metric in the presence of NLED instead of the original spacetime metric. However, we need to state that the metric describing the 4D ESTGB-NLED spacetime is obtained in the weak field where the NLED reduces to Maxwell's theory (see Ref. \cite{Canate:2020kla} for more detail). Therefore,  photons still follow the null geodesics of the original spacetime metric in the weak field.

Now, a general approach \cite{Bisnovatyi-Kogan:2010flt} is introduced to derive the deflection angle in the uniform or non-uniform plasma. We have the metric coefficients under the weak field approximation, which are given by
\begin{equation}\label{Q3}
g_{\alpha\beta}=\eta_{\alpha\beta} + h_{\alpha\beta},
\end{equation}
where $\eta_{\alpha\beta}$ is the Minkowski metric, i.e., $(-1,1,1,1)$, $h_{\alpha\beta}$  is perturbation metric. Note that
\begin{equation}\label{Q4}
h_{\alpha\beta} \ll 1, h_{\alpha\beta} \rightarrow 0  \quad {\rm where} \quad  x^{\alpha}\rightarrow \infty,
\end{equation}
\begin{equation}\label{Q04}
g^{\alpha\beta}=\eta^{\alpha\beta}-h^{\alpha\beta},~h^{\alpha\beta}=h_{\alpha\beta}.
\end{equation}

The refractive index of the  static inhomogeneous plasma that relies on the  photon frequency $ \omega(x^{i})  $ and space location $x^{\alpha}$ has the following form
\begin{equation}\label{Q10}
n^{2}=1-\frac{\omega^{2}_{e}}{\omega^{2}(x^{i})}, ~~\omega^{2}_{e}=\frac{4\pi e^{2} N(r)}{m}=K_{e} N(r),
\end{equation}
where $\omega_{e} $ is the electron plasma frequency, $N(r)$ is the electron density in the inhomogeneous plasma, $e$ and $m$ denote the charge and mass of the electron, respectively. It is worth noting that when  $\omega_{e}< \omega$  the electromagnetic waves can propagate in the such plasma. That is to say, the plasma medium has a reflective medium effect  when $\omega_{e}< \omega$ where $ \omega(\infty)\equiv \omega$.

Considering the effect of the plasma on the deflection angle in the weak field limit, we get the expression of deflection angle in the following form
\begin{equation}\label{Q17}
\hat\alpha_{k}=\frac{1}{2}\int_{-\infty}^{\infty}\bigg(h_{33,k}+ \frac{h_{00,k}}{1-\omega^{2}_{e}/\omega^{2}}-\frac{K_{e}N_{,k}}{\omega^{2}-\omega_{e}^{2}}\bigg)dz,
\end{equation}
for $k=1,2$. The deflection angle with the impact parameter $b$  found in Ref.\cite{Bisnovatyi-Kogan:2010flt} for more detail,  can be written as
\begin{equation}\label{Q18}
\hat\alpha_{k}=\frac{1}{2}\int_{-\infty}^{\infty}\frac{b}{r} \times
\bigg(\frac{dh_{33}}{dr}+\frac{1}{1-\omega^{2}_{e}/\omega^{2}}\frac{dh_{00}}{dr}-\frac{K_{e}}{\omega^{2}-\omega_{e}^{2}}\frac{dN}{dr}\bigg)dz.
\end{equation}

The location of the photon is presented by $b$ and $z$ under the axially symmetric case, and then the magnitude of the radius-vector is written as $r=\sqrt{b^{2}+z^{2}}$ \cite{Hensh:2019ipu}. It is worth noting that the negative value of $\hat\alpha_{b}$ indicates the bending of the photon trajectory towards the compact object, and the positive value indicates the opposite.

In the weak gravitational field regime, we can rewrite the metric around the 4D ESTGB  black hole as
\begin{equation}\label{Q19}
ds^{2}=ds_{0}^{2}+\bigg(\frac{R_{s}}{r}+\frac{q^{3}}{r^{3}}\bigg)(dt^{2}+dr^{2}),
\end{equation}
where $ds^{2}_{0} $ is the flat part of metric, and it has the following form
\begin{equation}\label{Q20}
ds^{2}_{0}=-dt^{2}+dr^{2}+r^{2}(d\theta^{2}+\sin^{2}\theta d\phi^{2}).
\end{equation}

The components $h_{\alpha\beta}$ can be expressed in the Cartesian frame as
\begin{equation}\label{Q21}
h_{00}=\frac{R_{s}}{r}+\frac{q^{3}}{r^{3}}, \quad h_{ik}=h_{00}n_{i}n_{k}, \quad
h_{33}=h_{00}\cos^{2}\chi,
\end{equation}
where $\cos \chi=z/\sqrt{b^{2}+z^{2}}$ and $r=\sqrt{b^{2}+z^{2}}$.

By substituting Eq.(\ref{Q21}) into Eq.(\ref{Q18}), we have the concrete form of the deflection angle in the following expression \cite{Bisnovatyi-Kogan:2010flt}
\begin{equation}\label{Q22}
\hat\alpha_{b}=\int_{-\infty}^{\infty}\frac{b}{2r}\bigg(\partial_{r}\big((\frac{R_{s}}{r}+\frac{q^{3}}{r^{3}})\cos^{2}\chi\big)
+\partial_{r}\big(\frac{R_{s}}{r}+\frac{q^{3}}{r^{3}}\big)\frac{1}{1-\omega^{2}_{e}/\omega^{2}}-\frac{K_{e}}{\omega^{2}-\omega^{2}_{e}}\partial_{r}N\bigg)dz.
\end{equation}

In what follows, we will calculate the integrals about the deflection angle considering the three specific plasma distributions, viz., uniform plasma, singular isothermal sphere (SIS), and non-singular isothermal sphere (NSIS) medium.

\subsection{Uniform plasma}
In the subsection, we will calculate the deflection angle using  Eq.(\ref{Q22}) for the photon propagating in  the 4D ESTGB spacetime surrounded by uniform plasma, which can be expressed as
\begin{equation}\label{Q23}
\hat\alpha_{\text{uni}}=\hat\alpha_{\text{uni1}}+\hat\alpha_{\text{uni2}}+\hat\alpha_{\text{uni3}}.
\end{equation}

The first term is the influence of the gravitational field of the ESTGB  black hole
\begin{equation}\label{Q24}
\hat\alpha_{\text{uni1}}=\int_{-\infty}^{\infty}\frac{b}{2r}\partial_{r}(\frac{R_{s}}{r^3}+\frac{q^{3}}{r^{5}})z^{2}dz = -\frac{R_{s}}{b}-\frac{2q^{3}}{3b^{3}}.
\end{equation}

Note that when $q=0$ the spacetime will recover to  the Schwarzschild spacetime, and we will obtain  $\hat\alpha_{\text{uni1}}=R_{s}/b$. The second term includes the influence of the  gravitational field and plasma medium,  which can be written as
\begin{equation}\label{Q25}
\hat\alpha_{\text{uni2}}=\int_{-\infty}^{\infty}\frac{b}{2r}\partial_{r}\big(\frac{R_{s}}{r}+\frac{q^{3}}{r^{3}}\big)\frac{1}{1-\omega^{2}_{e}/\omega^{2}}dz
=-(\frac{R_{s}}{b}+\frac{q^{3}}{b^{3}})\frac{1}{1-\omega^{2}_{e}/\omega^{2}}.
\end{equation}

Because the last term is the influence of the inhomogeneity of plasma, we get  $\partial_{r}N=0$ for uniform plasma.

In the relevant literature about weak gravitational lensing, the deflection angle is usually defined as a positive one \cite{Synge:1960ueh}.  Thus, we have the following expression  about the uniform plasma
\begin{equation}\label{Q26}
\begin{aligned}
\hat\alpha_{\text{uni}}=\frac{R_{s}}{b}+\frac{2q^{3}}{3b^{3}}+(\frac{R_{s}}{b}+\frac{2q^{3}}{b^{3}})\frac{1}{1-\omega^{2}_{0}/\omega^{2}},
\end{aligned}
\end{equation}
where $\omega_{0}$=$\omega_{e}(\infty)$.

\begin{figure}[htbh]
	\centering
	{\includegraphics[width=0.45\textwidth,height=0.32\textwidth]{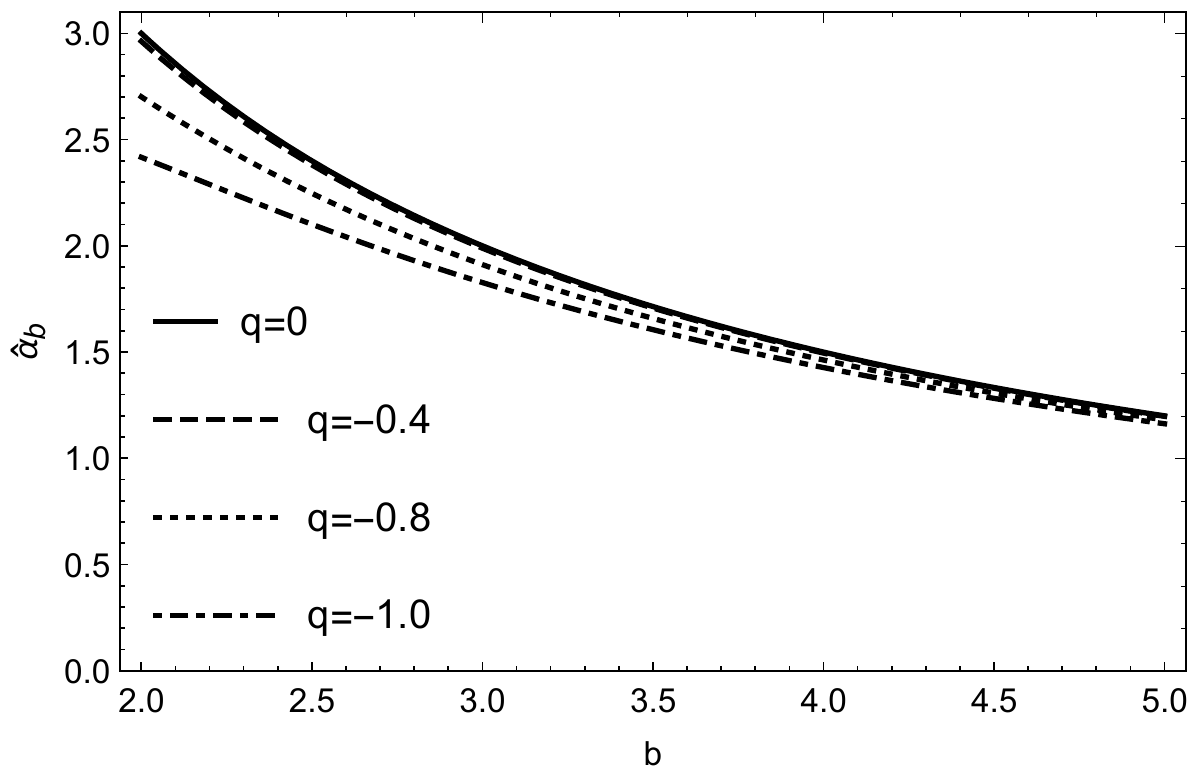}}
	{\includegraphics[width=0.45\textwidth,height=0.32\textwidth]{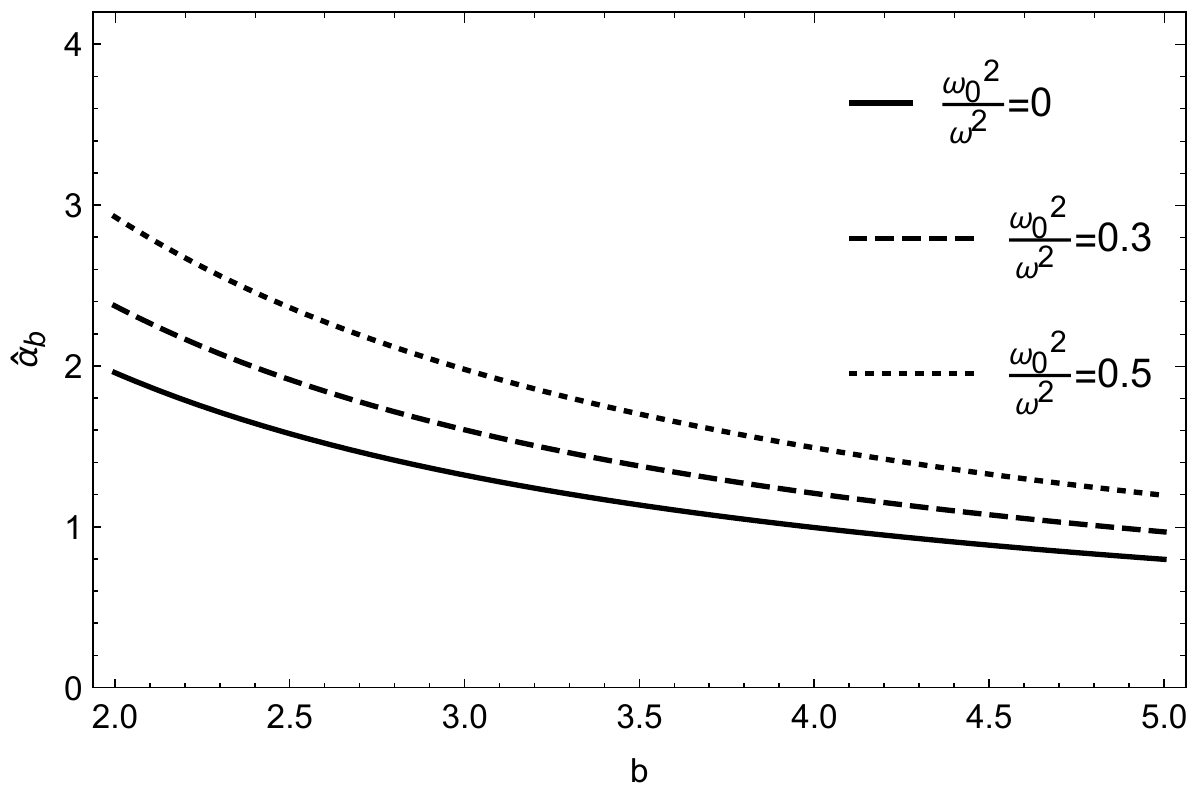}}
	\caption{The deflection angle $\hat\alpha_{b}$ as the function of impact parameter $b$ for different values of magnetic charge  at
		$\omega_{0}^{2}/\omega^{2}=0.5$ (left panel), and  uniform plasma medium parameter(right panel) at $q=-0.5$.}
	\label{fig1}
\end{figure}

\begin{figure}[htbh]
	\centering
	{\includegraphics[width=0.45\textwidth,height=0.32\textwidth]{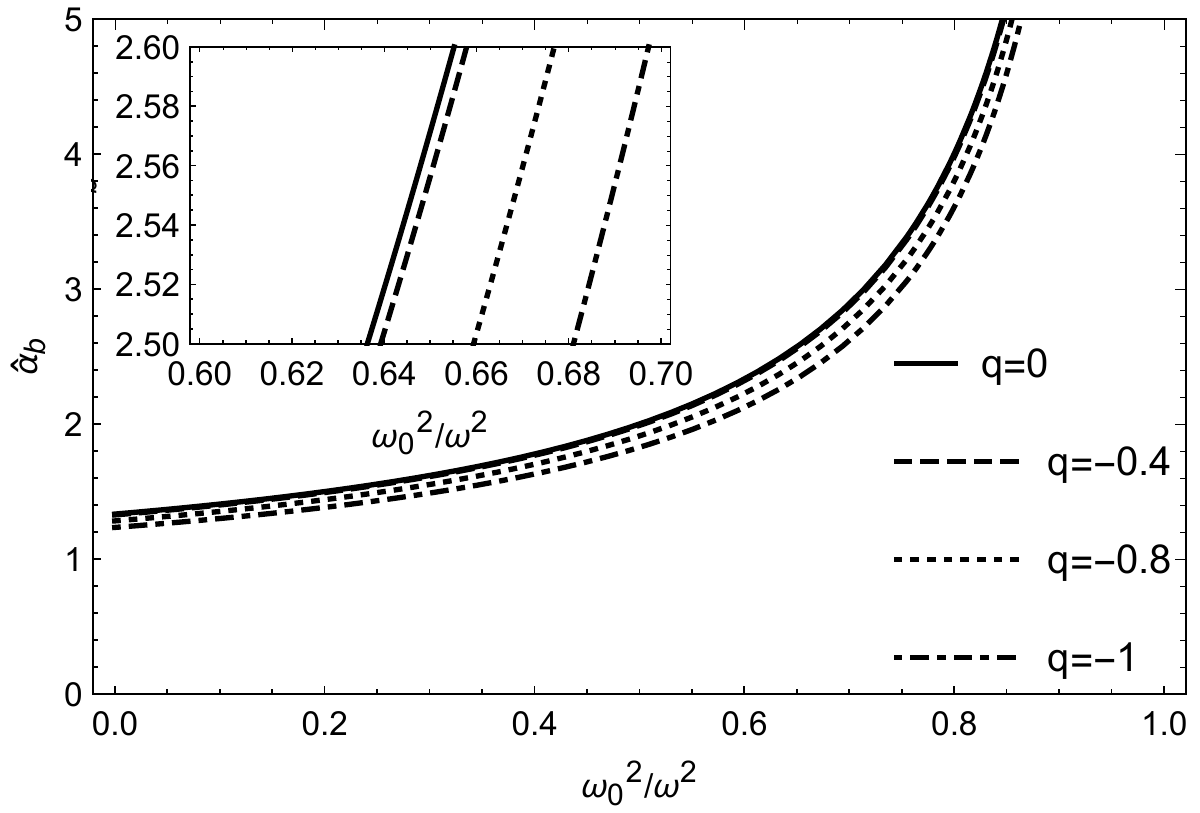}}
	{\includegraphics[width=0.45\textwidth,height=0.32\textwidth]{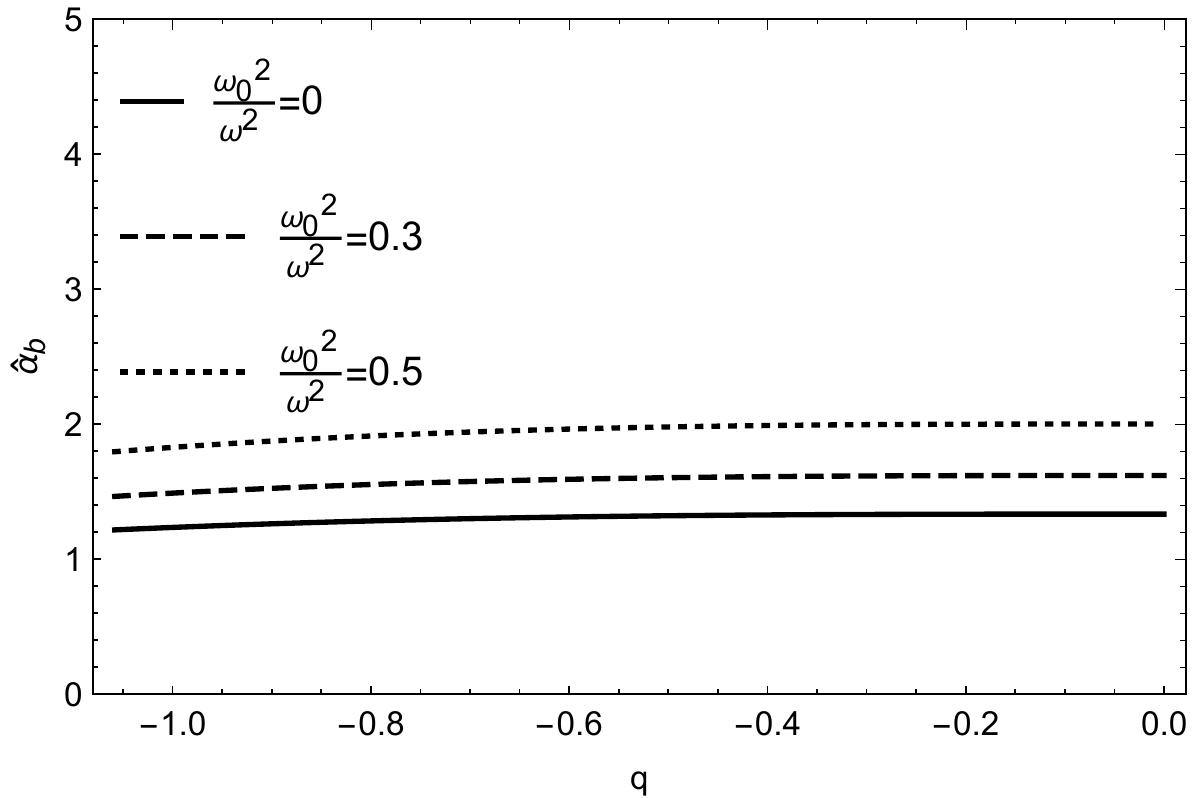}}
	\caption{The deflection angle $\hat\alpha_{b}$ as the function of uniform plasma medium parameter (left panel) and magnetic charge (right panel) for fixed $b=3$.}
	\label{fig2}
\end{figure}

In Fig.\ref{fig1}, we plot the deflection angle $\hat\alpha_{b}$ with respect to the impact parameter $b$ for different values of magnetic charge $q$ at $\omega_{0}^{2}/\omega^{2}=0.5$, and  plasma medium parameter at $q=-0.5$.  The deflection angle diminishes with an increase in the
impact parameter $b$.  As can be seen from Fig.\ref{fig1},  when $b\gg R_{s}$, we can neglect the effect of the magnetic charge on the deflection angle. In addition, it is easy to see from Eq.(\ref{Q26}), the deflection angle is very small or even disappear when the impact parameter $b$  is large. Fig.\ref{fig2} demonstrates the dependence of the deflection angle from the uniform plasma parameter and magnetic charge  at $b=3$. We can see in the left figure that the deflection angle increases rapidly  when $\omega_{0}^{2}/\omega^{2}$ increases to 1. As  the absolute value of magnetic charge decreases, the deflection angle slightly increases.

\subsection{Singular isothermal sphere}
In the subsection, we consider the case of an SIS around the 4D ESTGB  black hole. The SIS is primarily introduced in Refs.\cite{S1958} and \cite{J1987} to study the lens systems of the galaxies and clusters of galaxies. The  density distribution of the  SIS is written as
\begin{equation}\label{Q27}
\rho(r)=\frac{\sigma_{v}^{2}}{2 \pi r^{2}},
\end{equation}
where $v$ is the one-dimensional velocity dispersion. We can obtain the plasma concentration by making use of Eq.(\ref{Q27}) and the following relation
\begin{equation}\label{Q28}
N(r)=\frac{\rho(r)}{\kappa m_{p}},
\end{equation}
in which $\kappa$ is a coefficient which is related to the contribution of dark matter, called by 1D coefficient, and $m$ is the mass of proton. The plasma frequency has the expression
\begin{equation}\label{Q29}
\omega^{2}_{e}=K_{e}N(r)=\frac{K_{e} \sigma^{2}_{v}}{2\pi\kappa m_{p}}r^{-2}.
\end{equation}

Using Eq.(\ref{Q22}), we can calculate the deflection angle for an SIS. Due to the fact that the first term is  the effect of the gravitational field, it has the same expression as Eq.(\ref{Q24})
\begin{equation}\label{Q30}
\hat\alpha_{\text{sis1}}=\hat\alpha_{\text{uni1}}.
\end{equation}

For the other terms, we calculate  the integrals and obtain the following results
\begin{equation}\label{Q31}
\begin{aligned}
\hat\alpha_{\text{sis2}}&=\int_{-\infty}^{\infty}\frac{b}{2r}\partial_{r}\big(\frac{R_{s}}{r}+\frac{q^{3}}{r^{3}}\big)(1+\omega^{2}_{e}/\omega^{2})dz\\
&=-\big((\frac{R_{s}}{b}+\frac{2q^{3}}{b^{3}})+(\frac{2R_{s}}{3\pi b}+\frac{8q^{3}}{5\pi b^{3}})\frac{ \omega^{2}_{c}R_{s}^{2}}{\omega^{2} b^{2}}\big),
\end{aligned}
\end{equation}
\begin{equation}\label{Q031}
\hat\alpha_{\text{sis3}}=-\frac{K_{e}b}{2\omega^{2}}\int_{-\infty}^{\infty}\frac{1}{r}\frac{dN(r)}{dr}dz=\frac{\omega_{c}^{2}R_{s}^{2}}{2\omega^{2}b^{2}},
\end{equation}
where   $\omega_{c}^{2} $ is defined as \cite{Atamurotov:2021hoq}
\begin{equation}\label{Q32}
\omega_{c}^{2}=\frac{K_{e}\sigma^{2}}{2\kappa m_{p} R^{2}_{s}}.
\end{equation}

We obtain the deflection angle about the SIS,  which can be written as
\begin{equation}\label{Q33}
\hat\alpha_{\text{sis}}=\big((\frac{2R_{s}}{b}+\frac{8q^{3}}{3b^{3}})+(-\frac{1}{2}+\frac{2R_{s}}{3\pi b}+\frac{8q^{3}}{5\pi b^{3}})\frac{\omega_{c}^{2}R_{s}^{2}}{\omega^{2}b^{2}}\big).
\end{equation}

\begin{figure}[htbh]
	\centering
	{\includegraphics[width=0.45\textwidth,height=0.32\textwidth]{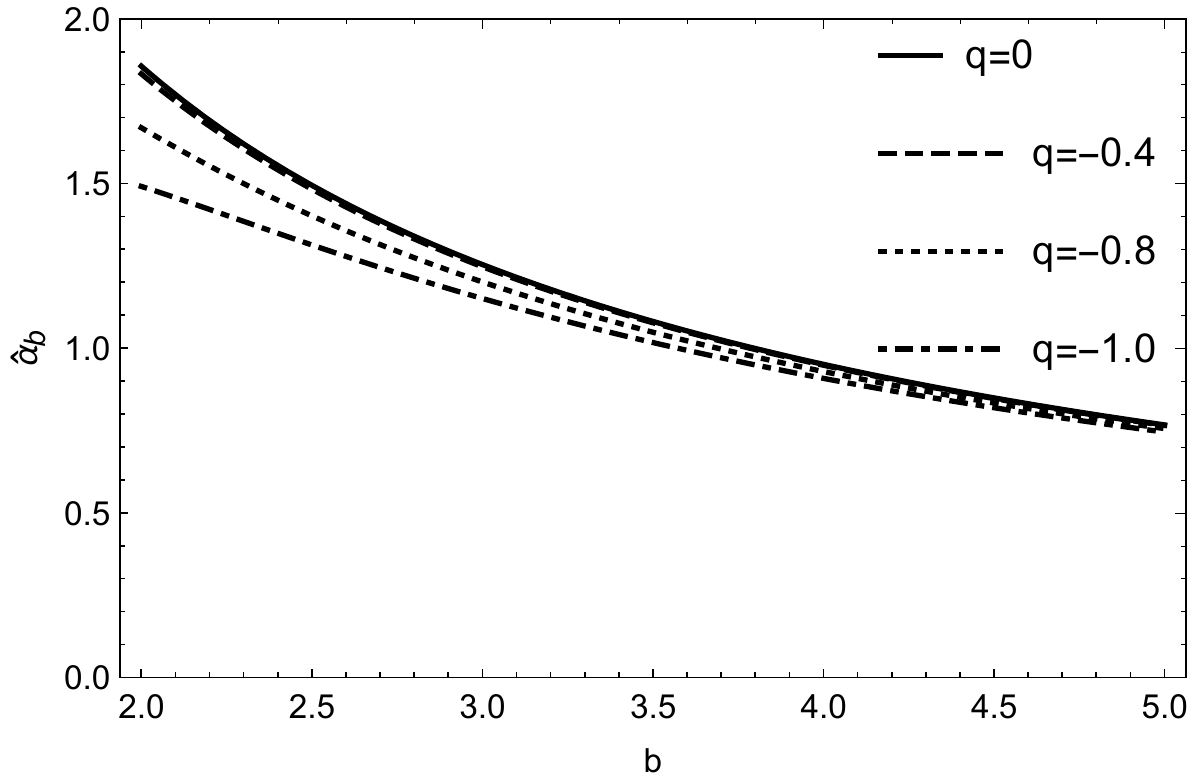}}
	{\includegraphics[width=0.45\textwidth,height=0.32\textwidth]{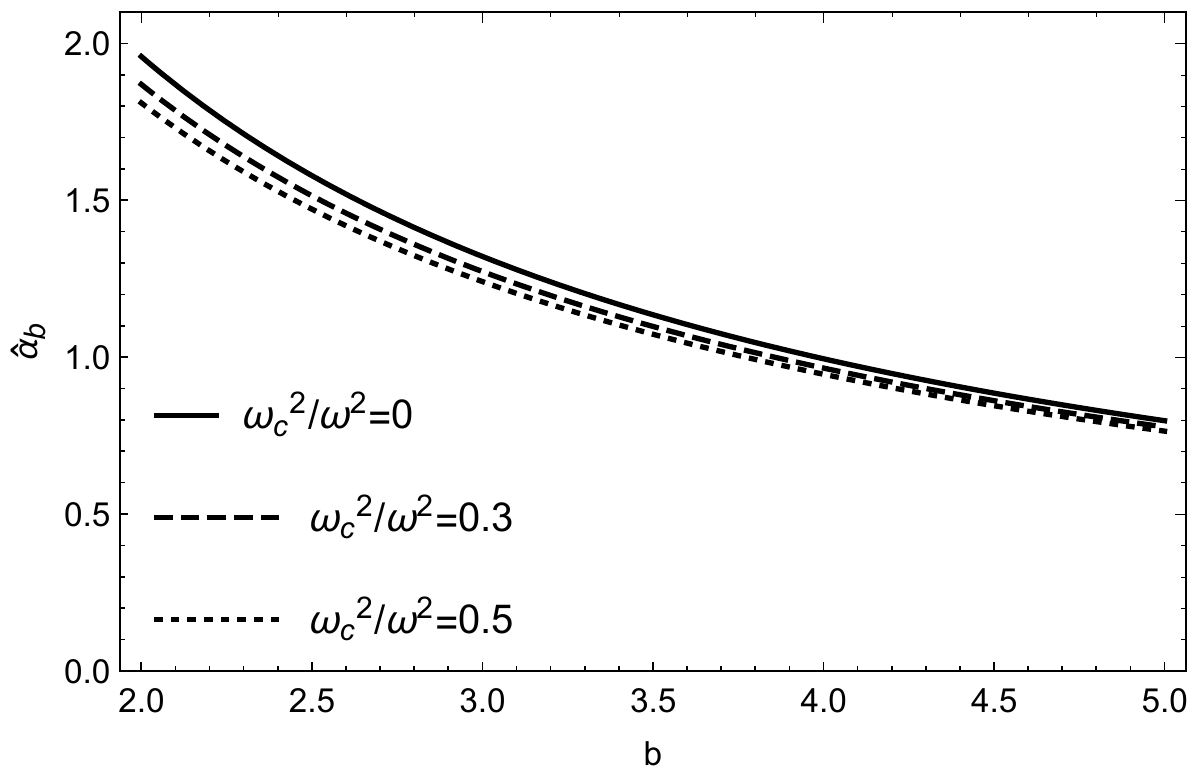}}
	\caption{The defection angle $\hat\alpha_{b}$ as the function of impact parameter $b$ for different values of magnetic charge (left panel) at  $\omega_{c}^{2}/\omega^{2}=0.5$, and SIS parameter(right panel) at $q=-0.5$.}
	\label{fig3}
\end{figure}

\begin{figure}[htbh]
	\centering
	{\includegraphics[width=0.45\textwidth,height=0.32\textwidth]{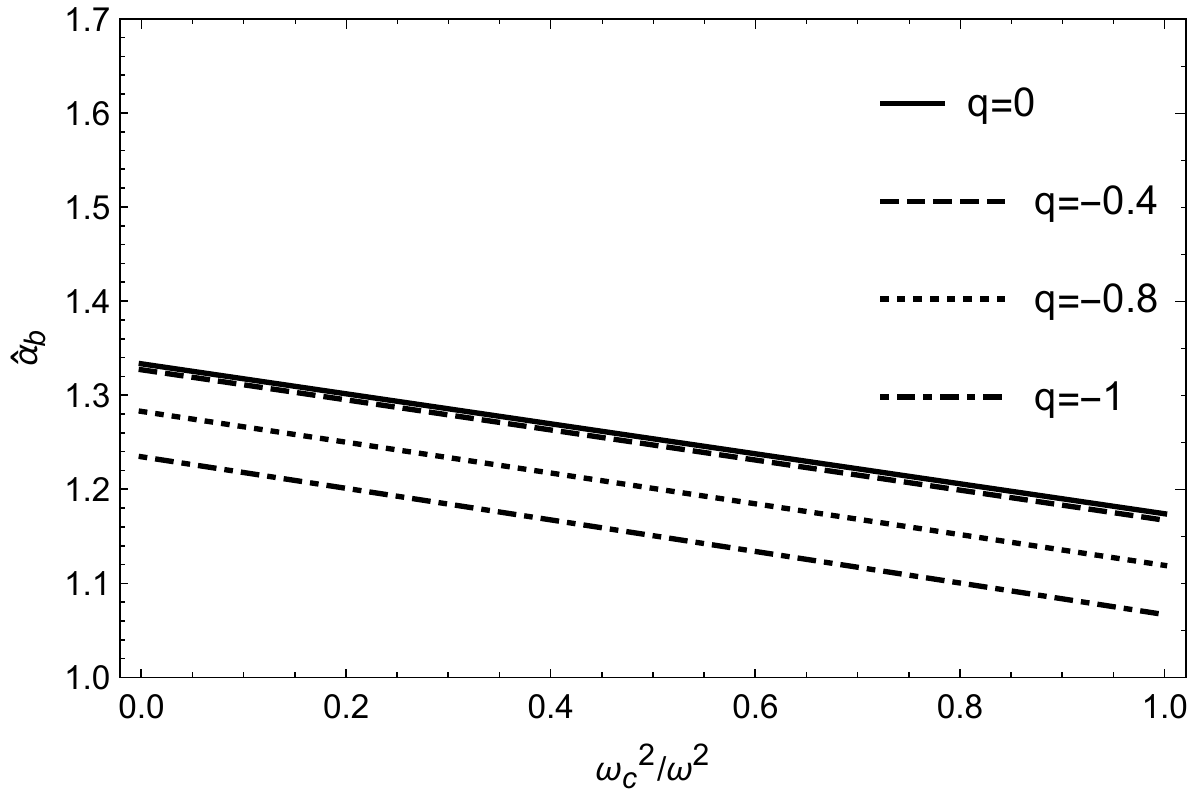}}
	{\includegraphics[width=0.45\textwidth,height=0.32\textwidth]{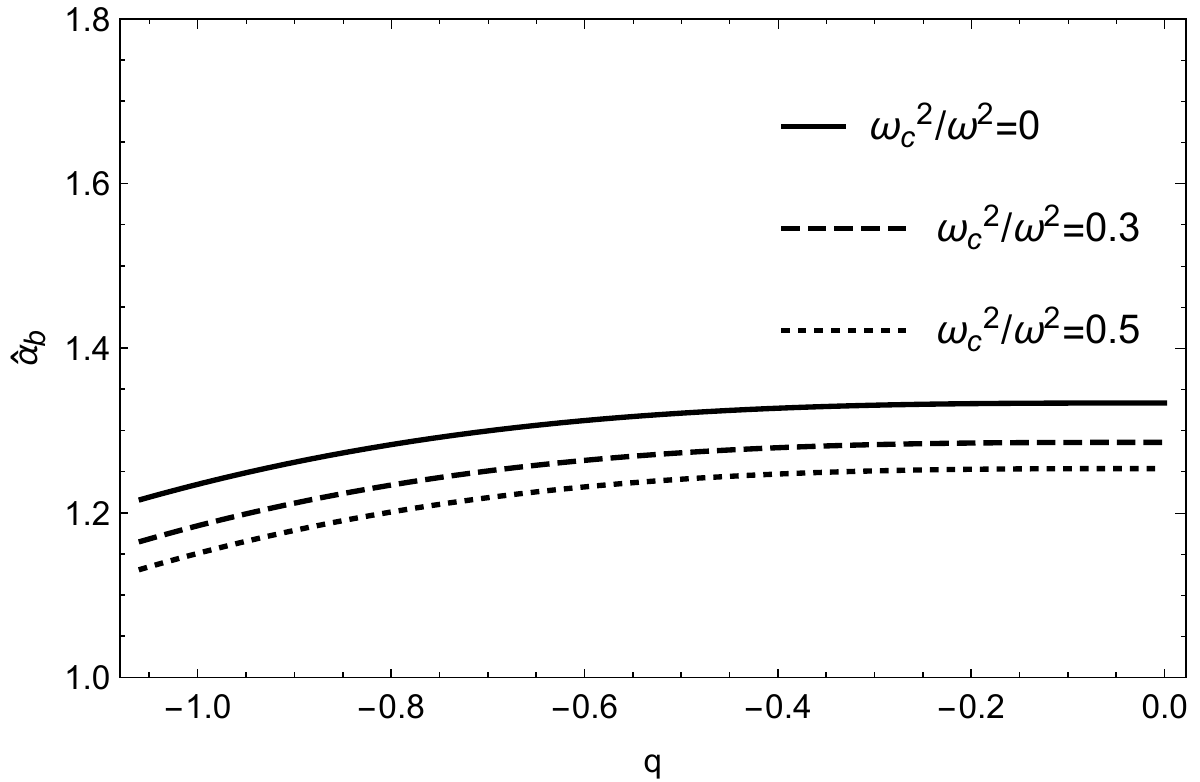}}
	\caption{The deflection angle $\hat\alpha_{b}$ as the function of SIS parameter(left panel) and magnetic charge (right panel) for fixed $b=3$.}
	\label{fig4}
\end{figure}

To simulate the effect of SIS on the trajectory of light, we demonstrate the deflection angle $\hat{\alpha}$ versus the impact parameter $b$ for different values of magnetic charge when $\omega_{c}^{2}/\omega^{2}$ is set to 0.5, and the SIS parameter for fixed $q=-0.5$ in Fig.\ref{fig3}.  It's not hard to get that when we increase the impact parameter the deflection angle decreases.  Fig.\ref{fig4} is the visualization of deflection angle to SIS parameter and magnetic charge, respectively.  It is straightforward to show that the deflection angle diminishes when $\omega_{c}^{2}/\omega^{2}$ increases (left figure), however, when the absolute value of  magnetic charge decreases the deflection angle increases (right figure). This means that the existence of a  SIS around the black hole reduces the deflection angle  in comparison to the vacuum or uniform cases.

\subsection{Non-singular isothermal  sphere}
In the subsection, we aim to give  the  exact expression of the  deflection angle of the ESTGB  black hole in  the presence of the NSIS. The plasma distribution can be expressed as \cite{Synge:1960ueh}
\begin{equation}\label{Q34}
\rho(r)=\frac{\sigma^{2}_{v}}{2\pi(r^{2}+r_{c}^{2})},
\end{equation}
where $r_{c}$ is the core radius, and the concentration becomes
\begin{equation}\label{Q35}
N(r)=\frac{\sigma^{2}}{2\pi \kappa m_{p}(r^{2}+r_{c}^{2})}.
\end{equation}

The corresponding plasma frequency has the following form
\begin{equation}\label{Q36}
\omega_{e}^{2}=\frac{K_{e} \sigma^{2}_{v}}{2\pi\kappa m_{p}(r^{2}+r_{c}^{2})}.
\end{equation}

Similarly to the last subsection, the first term remains unchanged, and other terms of Eq.(\ref{Q22}) will have the expressions
\begin{equation}\label{Q37}
\begin{aligned}
\hat\alpha_{\text{nsis2}}&=\int_{-\infty}^{\infty}\frac{b}{2r}\partial_{r}\big(\frac{R_{s}}{r}+\frac{q^{3}}{r^{3}}\big)(1+\omega^{2}_{e}/\omega^{2})dz\\
&=-(\frac{R_{s}}{b}+\frac{q^{3}}{b^{3}})-\bigg(\frac{R_{s}}{b\pi r_{c}^{2}}+\frac{b R_{s} \arctan{(\frac{r_{c}}{\sqrt{b^{2}+r_{c}^{2}}})}}{\pi r^{3}_{c}\sqrt{b^{2}+r_{c}^{2}}}\bigg)\\
&\times\frac{\omega_{c}^{2}R^{2}_{s}}{\omega^{2}}-\bigg(-\frac{1}{b^{2} r_{c}^{4}} +\frac{2}{3 b^{4} r_{c}^{2}} +\frac{\arctan{(\frac{r_{c}}{\sqrt{b^{2}+r_{c}^{2}}})}}{r_{c}^{5}\sqrt{b^{2}+r_{c}^{2}}}\bigg)\times\frac{3 q^{3} b R_{s}^{2} \omega_{c}^{2}}{\omega^{2} \pi},
\end{aligned}
\end{equation}
\begin{equation}\label{Q037}
\hat\alpha_{\text{nsis3}}=-\frac{K_{e}b}{2\omega^{2}}\int_{-\infty}^{\infty}\frac{1}{r}\frac{dN(r)}{dr}dz=\frac{b}{2(b^{2}+r_{c}^{2})^\frac{3}{2}} \frac{\omega_{c}^{2}R^{2}_{s}}{\omega^{2}},
\end{equation}
where
\begin{equation}\label{Q38}
\omega_{c}^{2}=\frac{K_{e}\sigma^{2}_{v}}{2\kappa m_{p} R^{2}_{s}}.
\end{equation}

One can obtain the  following form of the deflection angle by summing all the integrals
\begin{equation}\label{Q39}
\begin{aligned}
&\hat\alpha_{\text{nsis}}=(\frac{2R_{s}}{b}+\frac{8q^{3}}{3b^{3}})+\bigg(\frac{R_{s}}{b\pi r_{c}^{2}}-\frac{b}{2(b^{2}+r_{c}^{2})^\frac{3}{2}}+\frac{b R_{s} \arctan{(\frac{r_{c}}{\sqrt{b^{2}+r_{c}^{2}}})}}{\pi r^{3}_{c}\sqrt{b^{2}+r_{c}^{2}}}\bigg)\frac{\omega_{c}^{2}R^{2}_{s}}{\omega^{2}}\\
&~~~~~~+\bigg(-\frac{1}{b^{2} r_{c}^{4}} +\frac{2}{3 b^{4} r_{c}^{2}} \frac{\arctan{(\frac{r_{c}}{\sqrt{b^{2}+r_{c}^{2}}})}}{r_{c}^{5}\sqrt{b^{2}+r_{c}^{2}}}\bigg)\frac{3 q^{3} b R_{s}^{2} \omega_{c}^{2}}{\omega^{2} \pi}.
\end{aligned}
\end{equation}
\begin{figure}[htbh]
	\centering
	{\includegraphics[width=0.46\textwidth,height=0.32\textwidth]{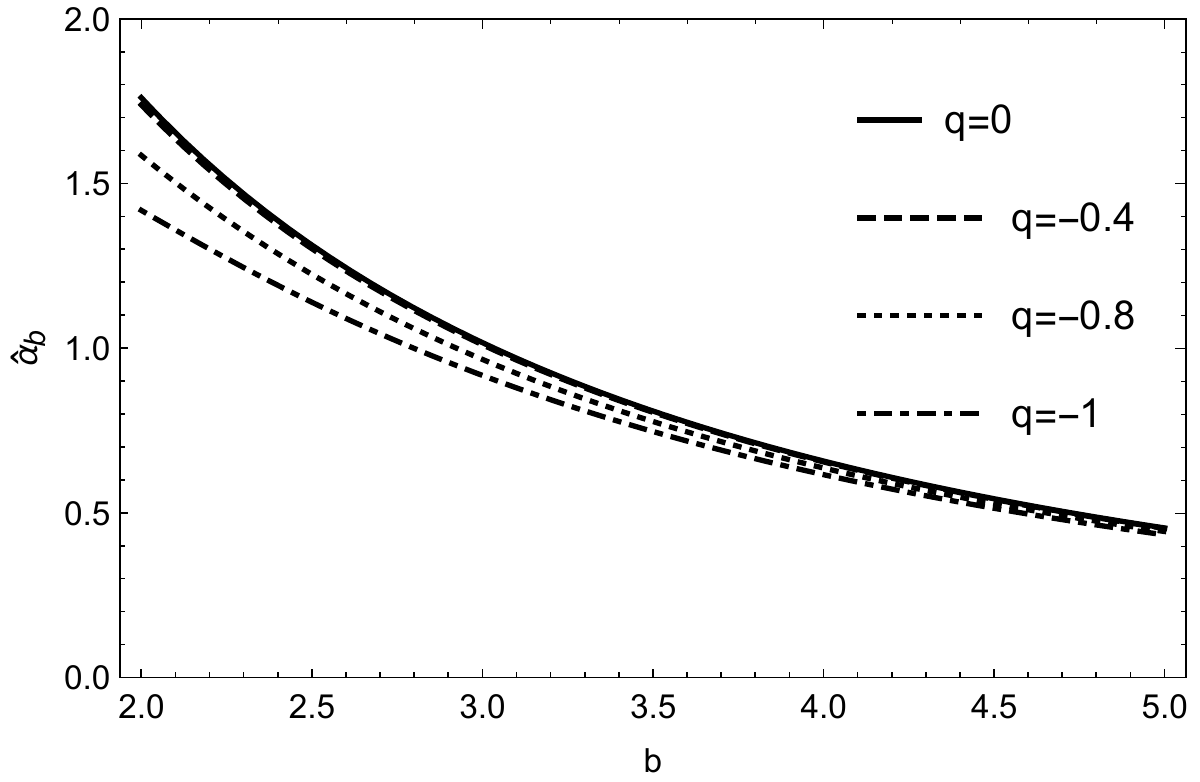}}
	{\includegraphics[width=0.46\textwidth,height=0.32\textwidth]{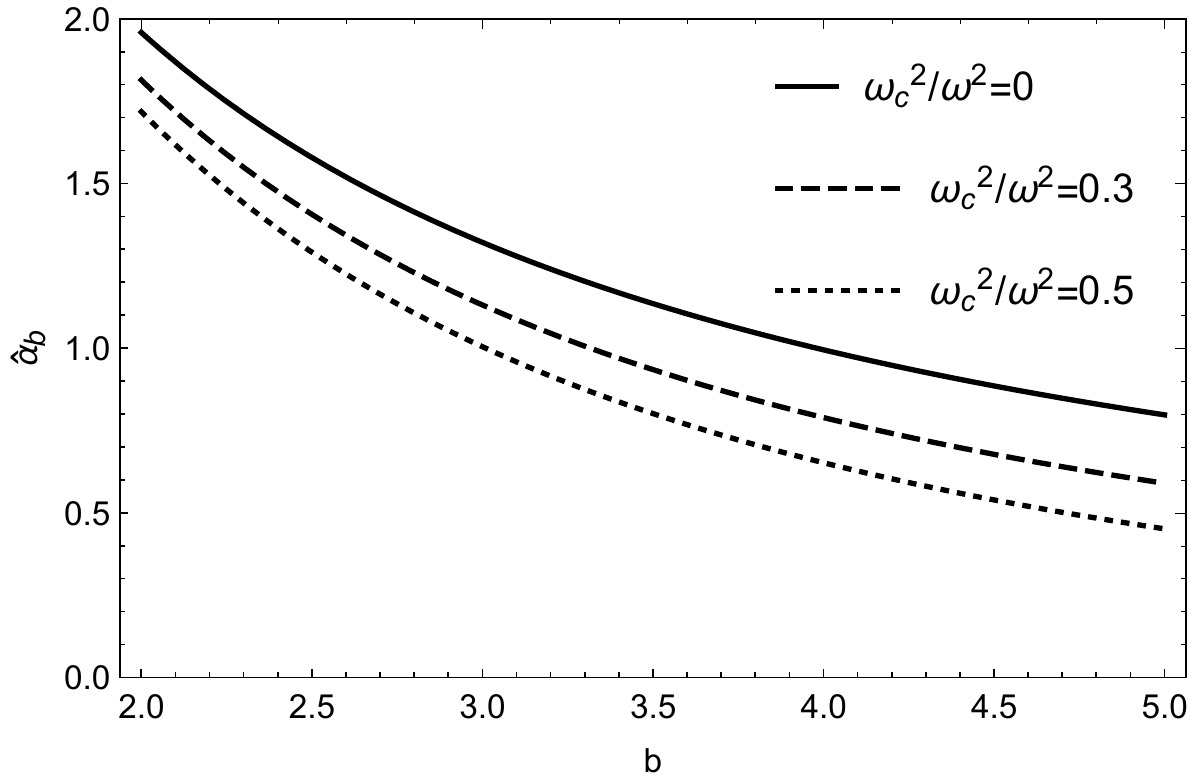}}
	\caption{The defection angle $\hat\alpha_{b}$ as the function of impact parameter $b$ for different values of magnetic charge (left panel) at $\omega_{c}^{2}/\omega^{2}=0.5$, and NSIS  parameter (right panel) at $q=-0.5$. For the case we take $r_{c}$=3.}
	\label{fig5}
\end{figure}

\begin{figure}[htbh]
	\centering
	{\includegraphics[width=0.46\textwidth,height=0.32\textwidth]{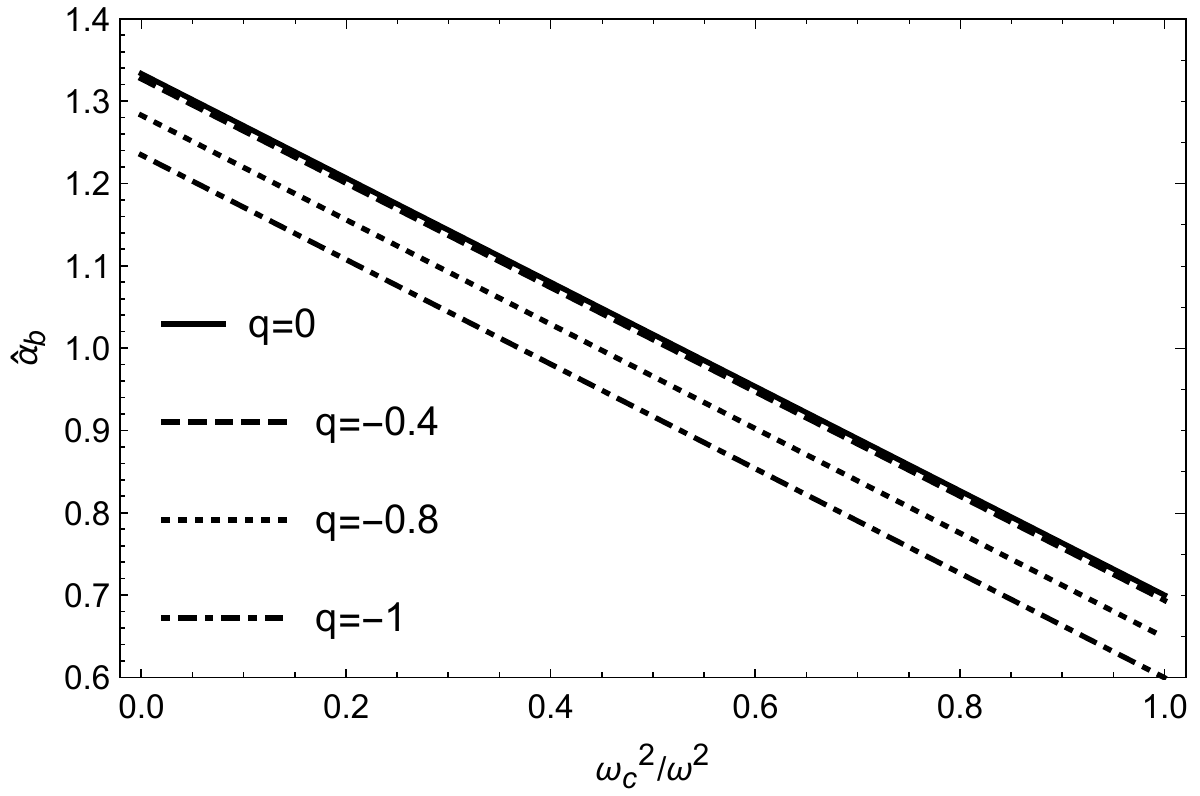}}
	{\includegraphics[width=0.46\textwidth,height=0.32\textwidth]{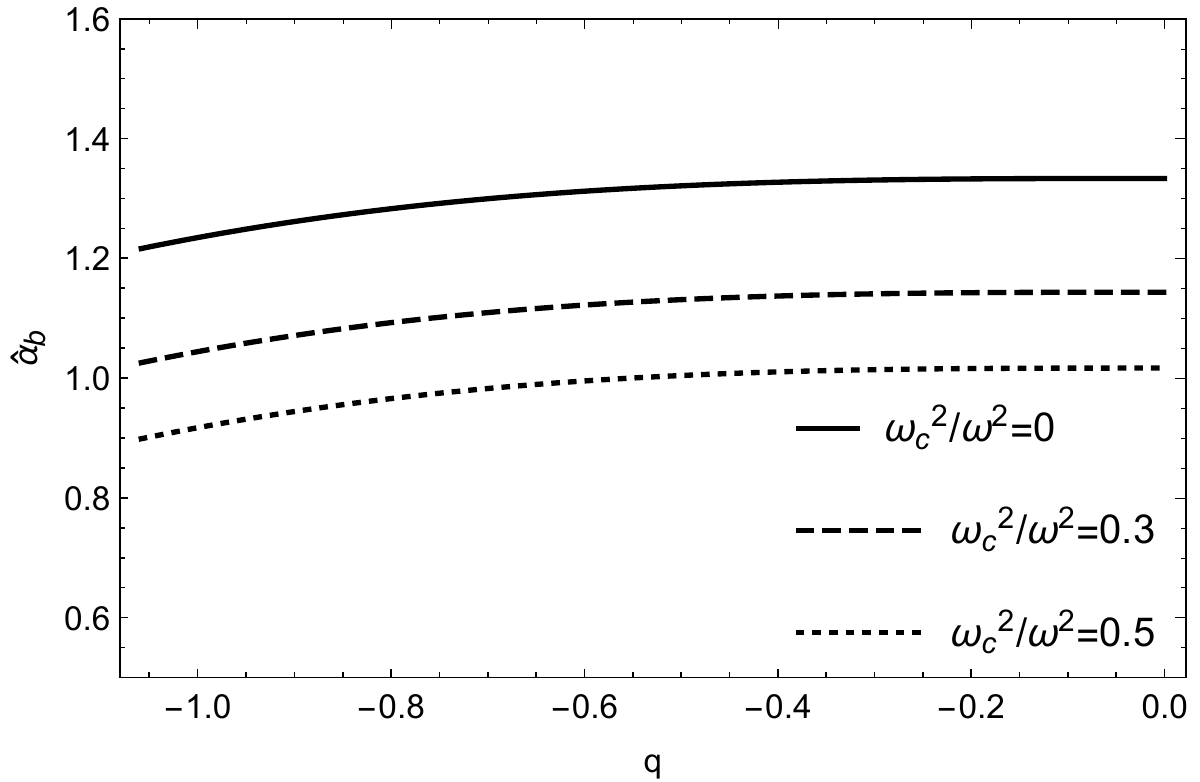}}
	\caption{The deflection angle $\hat\alpha_{b}$ as the function of  the NSIS  parameter (left panel) and magnetic charge (right panel)  for fixed $b=3$ and $r_{c}=3$.}
	\label{fig6}
\end{figure}

The variation of the deflection angle $\hat\alpha_{b}$ with the impact parameter $b$ is shown in Fig.\ref{fig5}, where the ESTGB-NLED black hole is surrounded by NSIS medium. From Fig.\ref{fig5}, we can conclude that the increase of the impact parameter leads to the diminishing of deflection angle. And we can see from the right panel that the difference in the deflection angle becomes more and more obvious with an increase in the impact parameter for the different values of the NSIS  medium. In Fig.\ref{fig6}, we plot the dependence of the deflection angle on the NSIS parameter for the different magnetic charges (left panel) and on the magnetic charge for the different NSIS parameters (right panel). In these two  cases we fix $b=3$ and $r_{c}=3$. The effect of NSIS on the deflection angle is similar to that of the SIS case by comparing Figs.\ref{fig4} and \ref{fig6}.

\begin{figure}[htbh]
	\centering
	{\includegraphics[width=0.44\textwidth,height=0.30\textwidth]{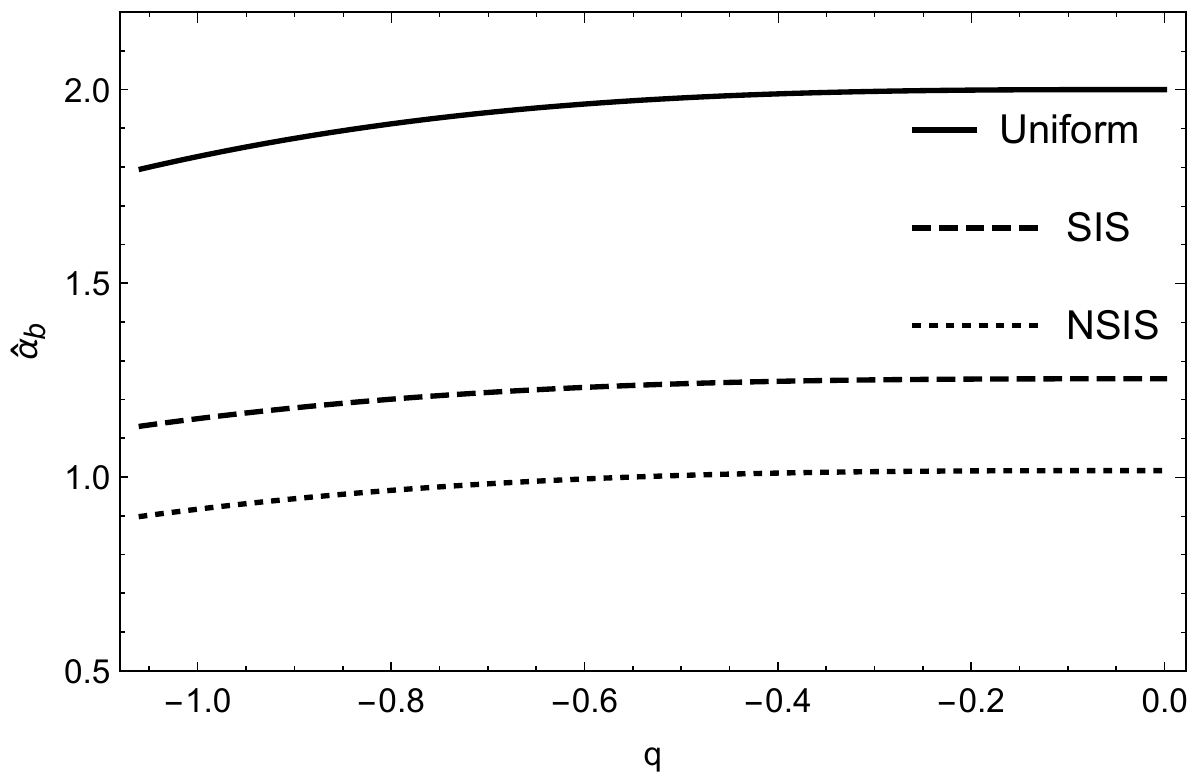}}
	{\includegraphics[width=0.44\textwidth,height=0.30\textwidth]{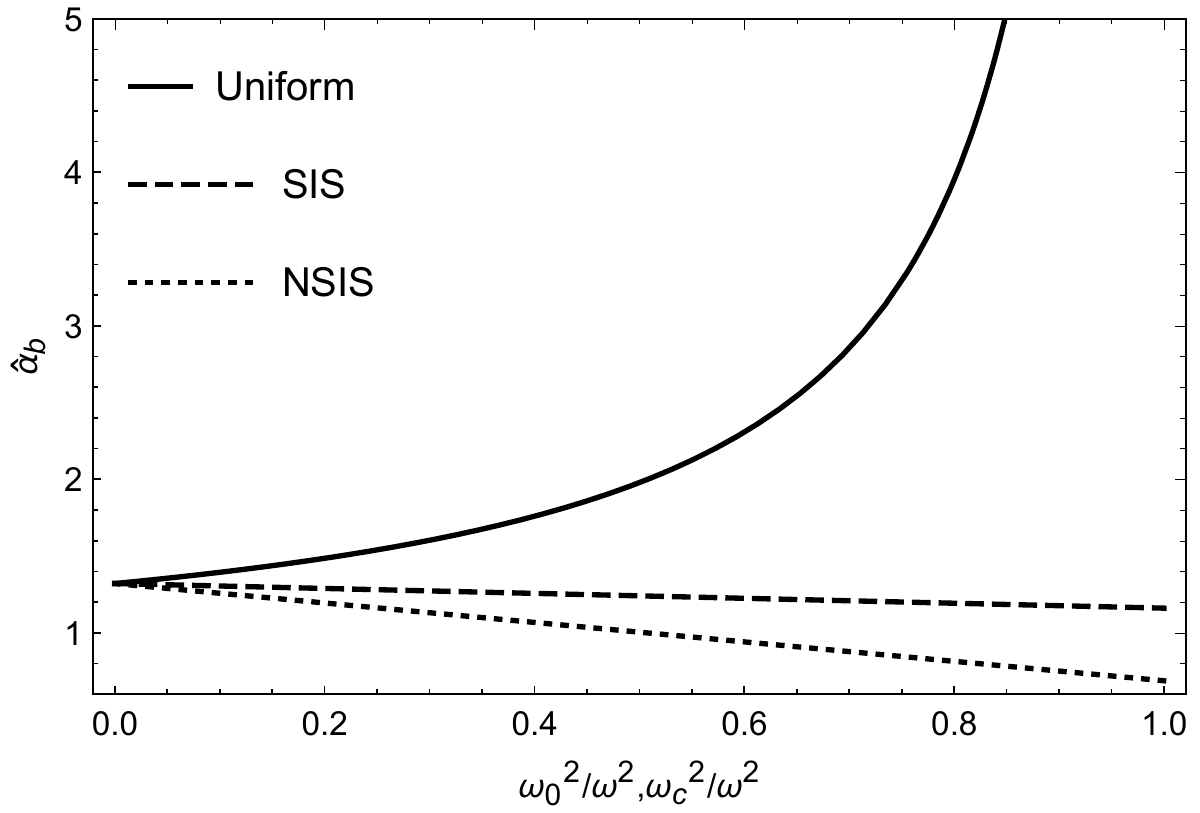}}
	{\includegraphics[width=0.45\textwidth,height=0.3\textwidth]{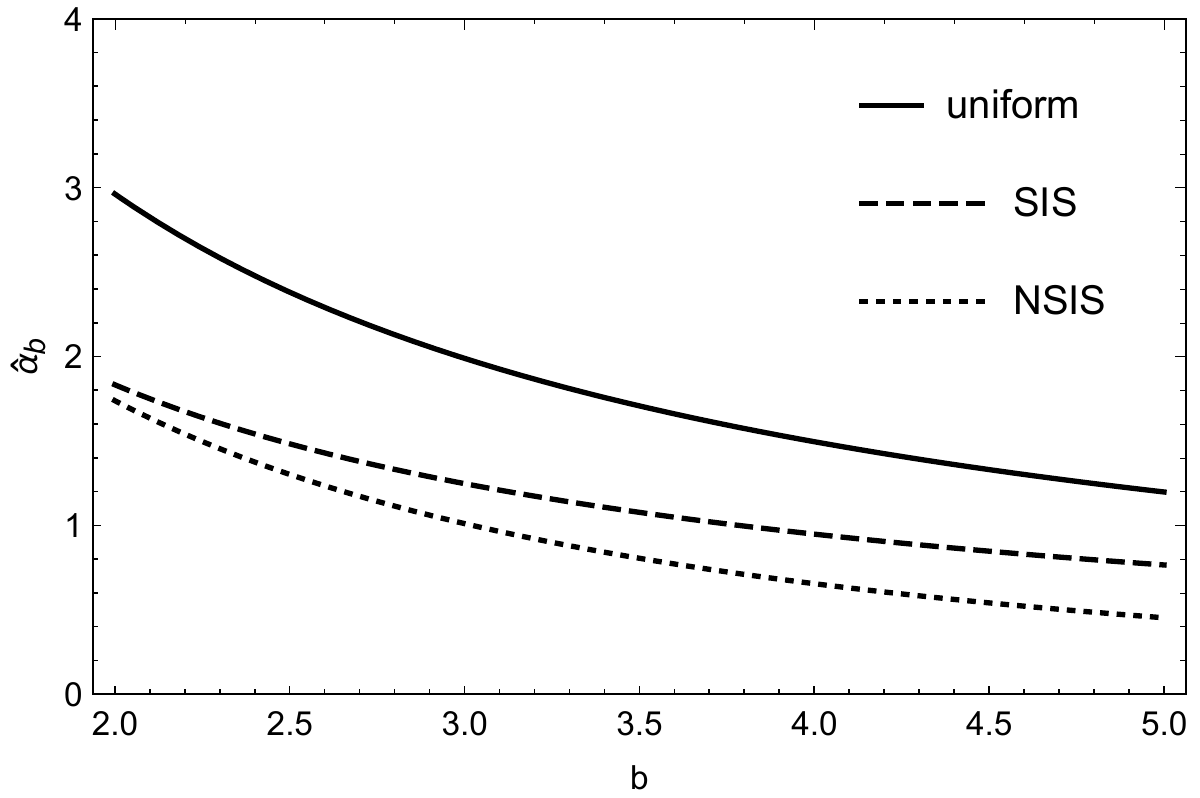}}
	\caption{The deflection angle $\hat\alpha_{b}$ as the function of  the magnetic charge (left panel), plasma parameter (right panel)  and impact parameter (bottom panel) for fixed $\omega_{0}^{2}/\omega^{2}=\omega_{c}^{2}/\omega^{2}=0.5$, $b=3$ and $r_{c}=3$.}
	\label{fig7}
\end{figure}

In the above three subsections, we studied the effect of the different distributions of the plasma and magnetic charge on the deflection angle in detail. To directly compare the effects of different plasmas, i.e., uniform plasma, SIS, and NSIS media, we study the dependence of the deflection angle on different parameters. The comparison results are shown in Fig.\ref{fig7} where we fix the corresponding parameters, viz., $\omega_{0}^{2}/\omega^{2}=\omega_{c}^{2}/\omega^{2}=0.5$, impact parameter $b=3$ and the core radius $r_{c}=3$.  The uniform plasma medium exhibits better refraction properties than the SIS and NSIS models, as shown in Fig.\ref{fig7}. It is easy to see that the magnetic charge has a small effect on the deflection angle of the black holes in different plasma distributions. We also notice from the  right figure that when we increase the plasma parameter $\omega_{0}^{2}/\omega^{2}$ or $\omega_{c}^{2}/\omega^{2}$, the deflection angle in the presence of the SIS or NSIS medium diminishes, whereas the deflection angle of the uniform plasma has the opposite trend. Finally, the deflection angle decreases with the increase of the impact parameter for  the three models. In a word, the bending degree of deflection can be expressed mathematically as, $\hat\alpha_{\text{uni}} > \hat\alpha_{\text{sis}}> \hat\alpha_{\text{nsis}}$.

\section{Magnification of image}\label{3}
\begin{figure}[htbh]
	\centering
	{\includegraphics[width=0.55\textwidth,height=0.40\textwidth]{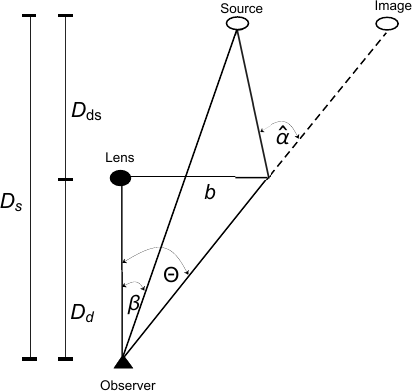}}
	\caption{Schematic diagram of gravitational lens effect }
	\label{fig07}
\end{figure}

In this section, we will analyze in detail the magnification of image for the ESTGB  black hole in the presence of the different plasma using the formula of the deflection angle studied in our previous section. The lens equation has the form \cite{Babar:2021exh}
\begin{equation}\label{Q40}
\theta D_{s}=\beta D_{s}+\hat\alpha_{b} D_{d s},
\end{equation}
where $ D_{s}$ is the distance from the observer to the distant light source, and $D_{d s}$ is the distance from the lens object to the distant light source (see Fig.\ref{fig07}). $\theta$  denotes the angle of the apparent source image for the observer lens axis,  $\beta$  denotes the angle of the light source with respect to the observer lens axis, and  $\hat\alpha_{b}$ is the angle between the apparent source image and light source, i.e., deflection angle. We make use of  the  relationship between  the impact parameter and angle $\theta$,  and $\theta$ possesses the expression $b=D_{d} \theta$  where $D_{d} $ is the distance from the lens object to the observer, to  rewrite the expression (\ref{Q40}), into the form \cite{Morozova}
\begin{equation}\label{Q41}
\beta =\theta-\frac{D_{ds}}{D_{s}}\frac{F(\theta)}{D_{d}}\frac{1}{\theta},
\end{equation}
and
\begin{equation}\label{Q42}
F(\theta)=|\hat\alpha_{b}|b=|\hat\alpha_{b}(\theta)|D_{d} \theta.
\end{equation}

Note that when the light source, lens object, and observer remain in a straight line, the angle $\beta$ is equal to zero. In such a case, the relativistic  image will form a  relativistic ring known as an Einstein ring. The radius of the Einstein ring $R_{0}=D_{d} \theta_{0}$, where $ \theta_{0}$ denotes the Einstein angle. The Einstein angle in the context of the Schwarzschild black hole can be expressed as \cite{Morozova}
\begin{equation}\label{Q43}
\theta_{0}=\sqrt{2R_{s}\frac{D_{ds}}{D_{d}D_{s}}}.
\end{equation}

The Einstein angle $ \theta_{0}$  is small but can be solved with modern telescopes. However, we can detect the gravitational lensing owing to the changes in the apparent brightness of the source, namely magnification of the image brightness. The basic equation of the magnification of the image brightness is expressed as \cite{Morozova}
\begin{equation}\label{Q44}
\mu_{\Sigma}=\frac{I_{tot}}{I_{*}}=\sum\limits_{k}\bigg|\bigg(\frac{\theta_{k}}{\beta}\bigg)\bigg(\frac{d\theta_{k}}{d\beta}\bigg)\bigg|,~~k=1,2,...,s,
\end{equation}
where $I_{tot}$ and $I_{*}$ refer to the total brightness of the image and unlensed brightness of the pure source, respectively. $k$ is the number of the images and $s$ is the total number of the images.

Next, we will study the effect of the different distribution plasma around the ESTGB black hole on the magnification of the images.

\subsection{Uniform plasma}
We first  calculate the expression of  the Einstein angle $\theta^{pl}_{0} $  in the context of the uniform plasma. We have the form by using   Eqs.(\ref{Q26}) and (\ref{Q41}) as follows
\begin{equation}\label{Q45}
(\theta^{pl}_{0})_\text{uni}=\theta_{0}\bigg\{\frac{1}{2}\bigg((1+\frac{2q^{3}}{3R_{s}b^{2}})+(1+\frac{2q^{3}}{R_{s} b^{2}})\frac{1}{1-\omega_{0}^{2}/\omega^{2}}\bigg)\bigg\}^{\frac{1}{2}}.
\end{equation}

We obtain the magnification of image by bring the above Eq.(\ref{Q45}) into Eq.(\ref{Q44}), which is given by \cite{Morozova}
\begin{equation}\label{Q46}
\begin{aligned}
\mu_{tot}^{pl}=\mu_{+}^{pl}+\mu_{-}^{pl}=\frac{x^{2}+2}{x\sqrt{x^{2}+4}}.
\end{aligned}
\end{equation}

Here $\mu_{+} $ is the  magnification factor of the primary image, which is  located  on the same side of the light source with respect to the lens object \cite{Morozova}
\begin{equation}\label{Q47}
\mu_{+}=\frac{1}{4}\bigg[\frac{x}{\sqrt{x^{2}+4}}+\frac{\sqrt{x^{2}+4}}{x}+2\bigg],
\end{equation}
and $\mu_{-} $ is the  magnification factor  of the secondary image, which is  situated on the opposite side
\begin{equation}\label{Q48}
\mu_{-}=\frac{1}{4}\bigg[\frac{x}{\sqrt{x^{2}+4}}+\frac{\sqrt{x^{2}+4}}{x}-2\bigg],
\end{equation}
where $x$ denotes the dimensionless parameter in the presence of the uniform plasma. It has the following form
\begin{equation}\label{Q49}
x_\text{uni}=\frac{\beta}{(\theta^{pl}_{0})_\text{uni}}=x_{0}\bigg\{\frac{1}{2}\bigg((1+\frac{2q^{3}}{3R_{s}b^{2}})+(1+\frac{2q^{3}}{R_{s} b^{2}})\frac{1}{1-\omega_{0}^{2}/\omega^{2}}\bigg)\bigg\}^{-\frac{1}{2}},
\end{equation}
with $x_{0}=\beta/\theta_{0}$.

\begin{figure}[htbh]
	\centering
	{\includegraphics[width=0.46\textwidth,height=0.32\textwidth]{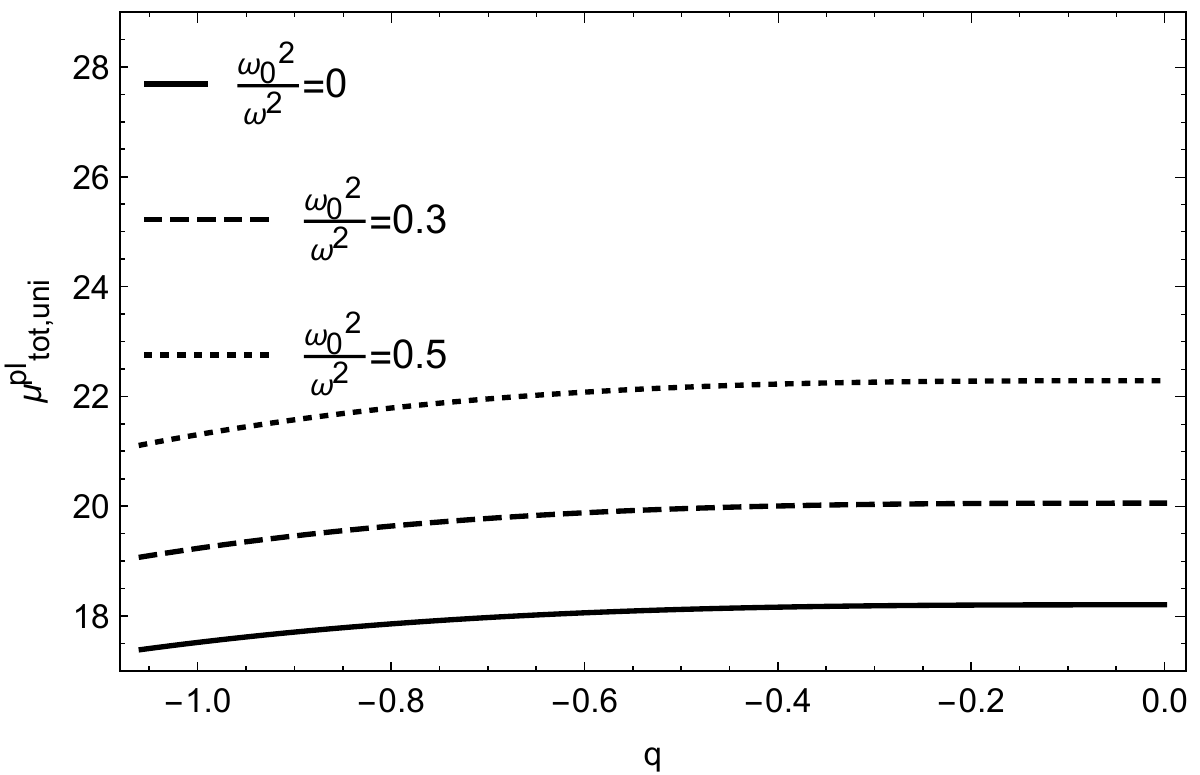}}
	{\includegraphics[width=0.46\textwidth,height=0.32\textwidth]{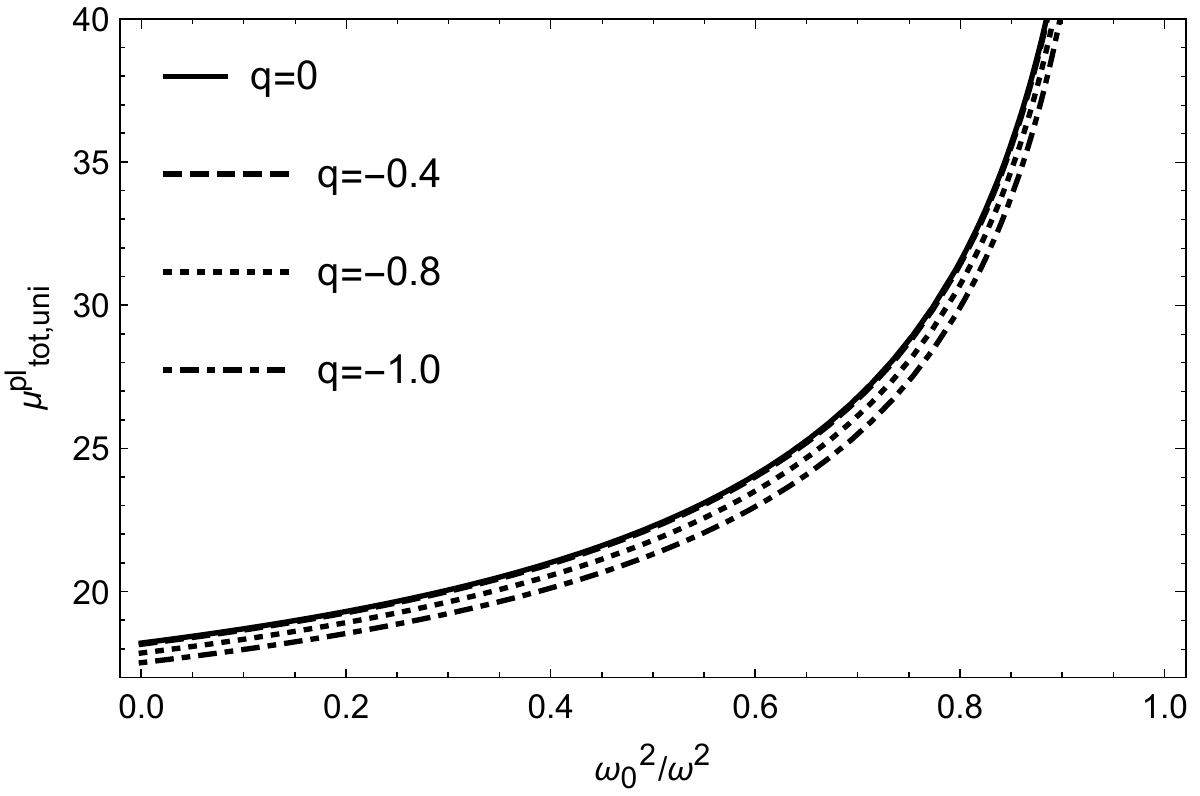}}
	\caption{The  total magnification of the image $\mu_{tot}$ as the function of  the magnetic charge (left panel), uniform plasma parameter (right panel)  with  fixed $R_{s}=2$, $b=3$ and $x_{0}=0.055$.}
	\label{fig9}
\end{figure}
\begin{figure}[htbh]
	\centering
	{\includegraphics[width=0.46\textwidth,height=0.32\textwidth]{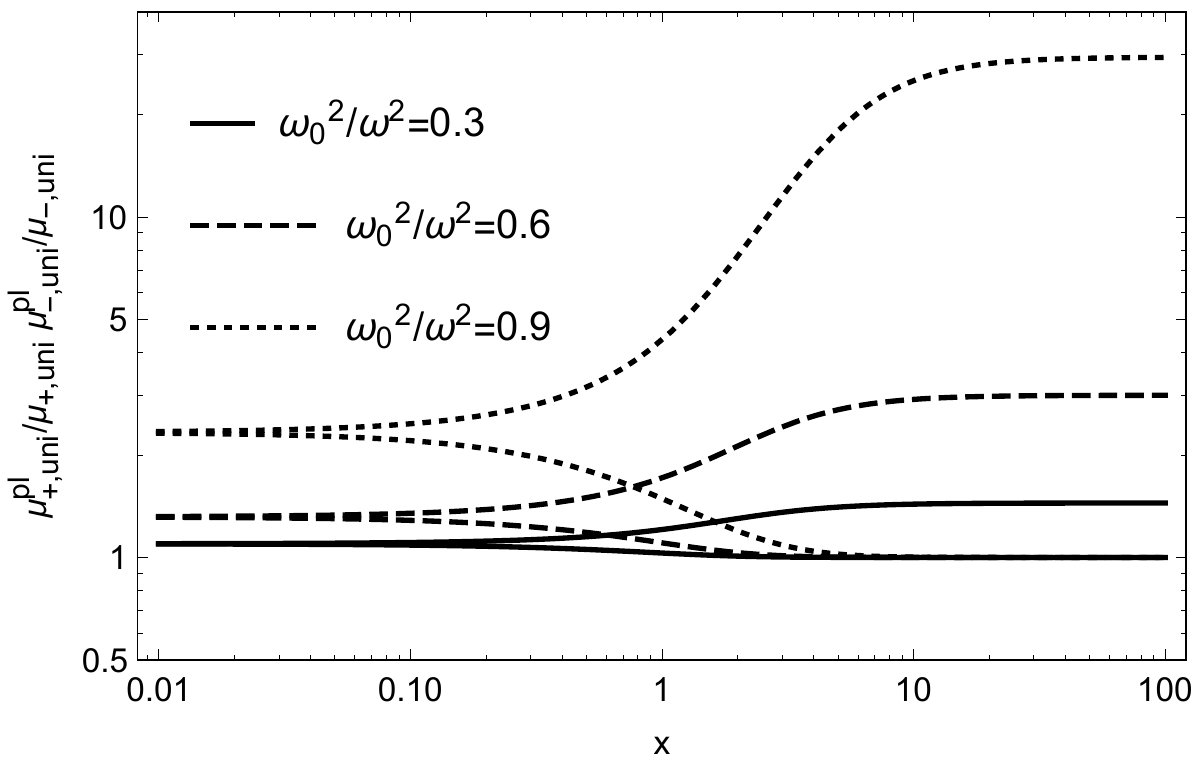}}
	\caption{ The ratio $\mu_{+}^\text{pl}/\mu_{+}$ (lower curves) and $\mu_{-}^\text{pl}/\mu_{-}$ (upper curves) of the magnification as the function of $x$  for the different values of the  uniform plasma distribution. We fix the  parameters as $q=-0.5$, $b=3$ and $R_{s}=2$.}
	\label{fig10}
\end{figure}

For a better understanding of the effect of  the magnetic charge  and plasma on the   magnification of image, in Fig.\ref{fig9}, we plot the variation of the total magnification of image with the magnetic charge for the different values of the uniform plasma parameter (left figure) and the uniform plasma parameter for the different values of magnetic charge (right figure) for fixed $R_{s}=2$, $b=3$ and $x_{0}=0.055$. We can see that the total magnification exhibits a small increase as the absolute value of magnetic charge decreases  and reaches a maximum when it returns to the Schwarzschild black hole. It is easy to see from the right panel that the total magnification increases exponentially with the increase of uniform plasma distribution. In other words, the existence of uniform plasma usually increases the magnification. Besides, we also plot  the ratios $\mu_{+}^\text{pl}/\mu_{+}$ (lower curves) and $\mu_{-}^\text{pl}/\mu_{-}$ (upper curves) of the magnification with the given parameters  $q=-0.5$, $b=3$ and $R_{s}=2$ in Fig.\ref{fig10}, for more details about the effect of the plasma on the magnification. It is evident that when the value of the uniform plasma density distribution increases, the magnification ratio increases. The behavior of the magnification ratio of the image brightness corresponds to the fact that the deflection angle is increased by $\omega_{0}^{2}/\omega^{2}$.  In addition, the magnification ratio of the secondary image $\mu_{-}^\text{pl}/\mu_{-}$ becomes larger, while the magnification ratio of the primary image $\mu_{+}^\text{pl}/\mu_{+}$ tends to unity  when $x$ increases.

\subsection{Singular isothermal sphere}
We have calculated the deflection angle for the case that the 4D ESTGB   black hole  surrounded by the uniform plasma in the last subsection. So in the subsection, we consider the influence of the SIS on the total magnification and the magnification ratio of image brightness.  The  expression of  the Einstein angle $\theta^{pl}_{0} $  in the context of the SIS medium can be expressed as
\begin{equation}\label{Q50}
(\theta^{pl}_{0})_\text{sis}=\theta_{0}\bigg\{\frac{1}{2}\bigg((2+\frac{8q^{3}}{3R_{s}b^{2}})+(-\frac{1}{2}+\frac{2R_{s}}{3\pi b}+\frac{8q^{3}}{ 5\pi b^{3}}) \frac{ R_{s} \omega_{c}^{2}}{b \omega^{2} }\bigg)\bigg\}^{\frac{1}{2}}.
\end{equation}

Since the calculational part is similar, we have $x$ in the presence of the SIS plasma medium, which has the following form
\begin{equation}\label{Q51}
x_\text{sis}=\frac{\beta}{(\theta^{pl}_{0})_\text{sis}}=x_{0}\bigg\{\frac{1}{2}\bigg((2+\frac{8q^{3}}{3R_{s}b^{2}})+(-\frac{1}{2}+\frac{2R_{s}}{3\pi b}+\frac{8q^{3}}{ 5\pi b^{3}}) \frac{ R_{s} \omega_{c}^{2}}{b \omega^{2} }\bigg)\bigg\}^{-\frac{1}{2}},
\end{equation}
where $x_{0}=\beta/\theta_{0}$.

\begin{figure}[htbh]
	\centering
	{\includegraphics[width=0.46\textwidth,height=0.32\textwidth]{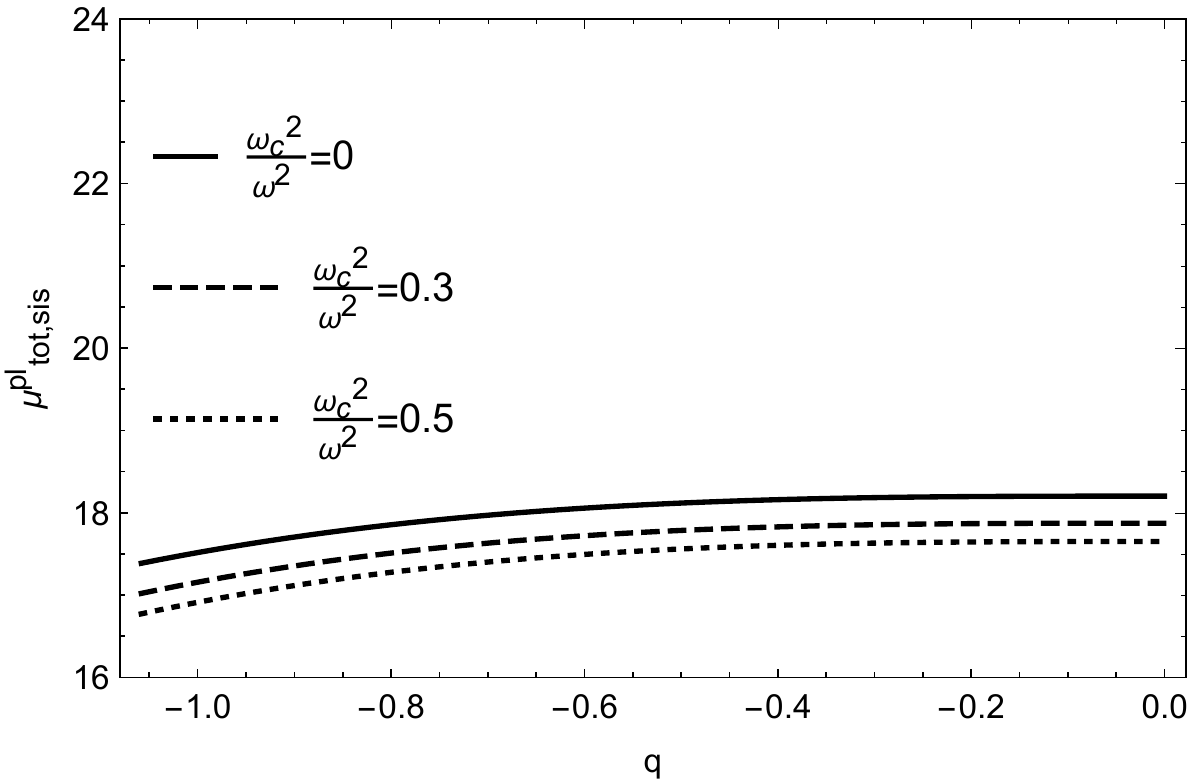}}
	{\includegraphics[width=0.46\textwidth,height=0.32\textwidth]{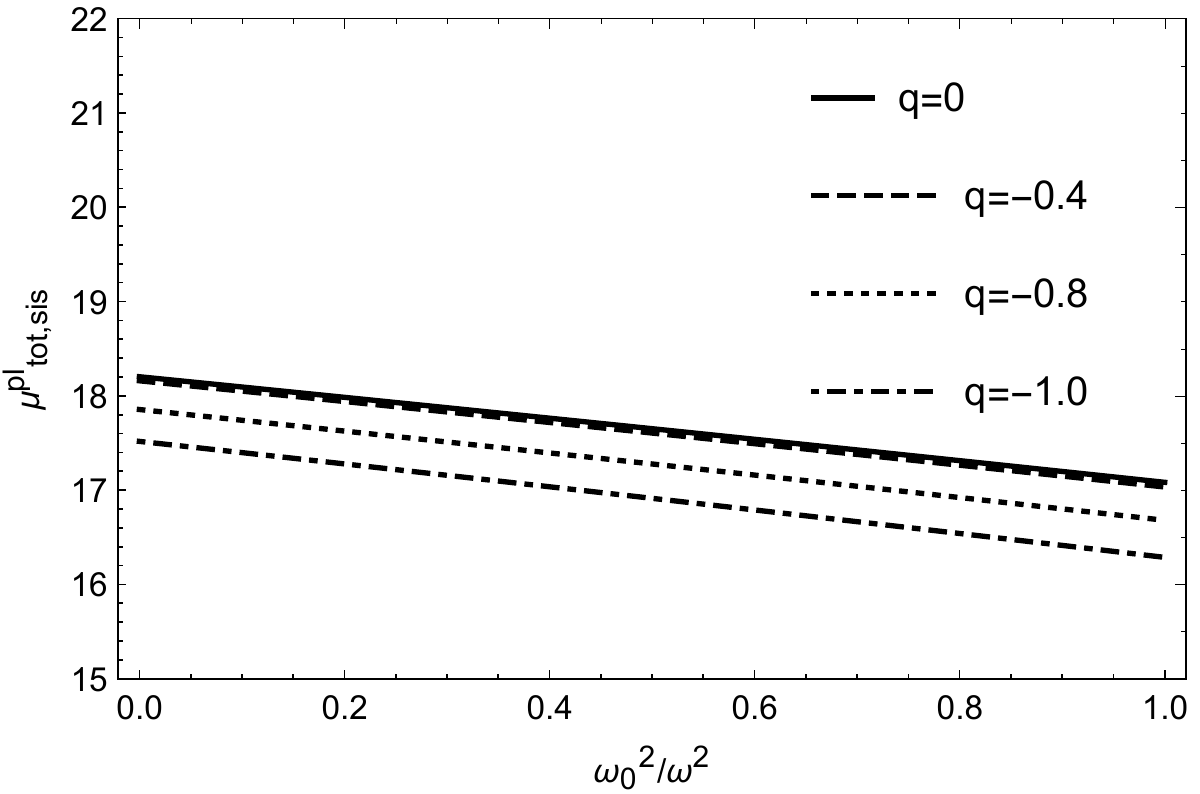}}
	\caption{The  total magnification of the image  brightness $\mu_{tot}$ as the function of  the
		magnetic charge (left panel), SIS parameter (right panel)  with  fixed $R_{s}=2$, $b=3$ and $x_{0}=0.055$.}
	\label{fig11}
\end{figure}

\begin{figure}[htbh]
	\centering
	{\includegraphics[width=0.46\textwidth,height=0.32\textwidth]{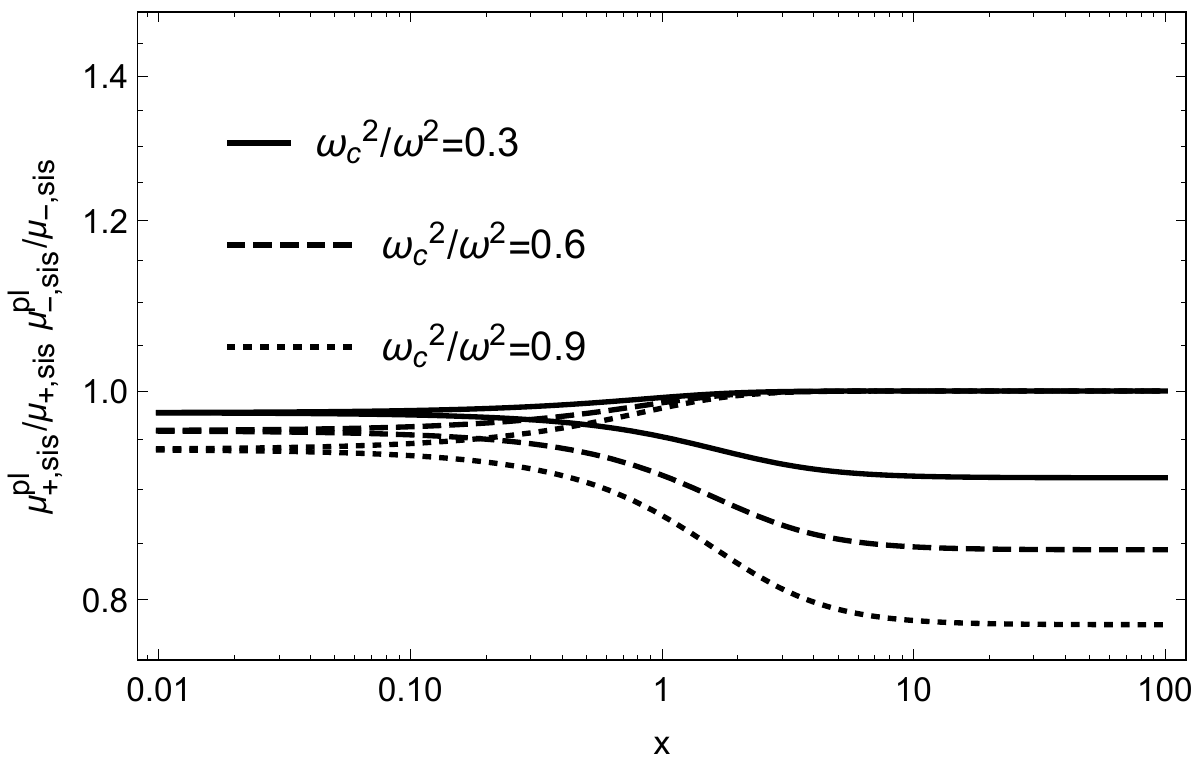}}
	\caption{ The ratio $\mu_{+}^\text{pl}/\mu_{+}$ (lower curves) and $\mu_{-}^\text{pl}/\mu_{-}$ (upper curves) of the magnification as the function of $x$  for the different values of the SIS parameter. We fix the  parameters as $q=-0.5$, $b=3$ and $R_{s}=2$.}
	\label{fig12}
\end{figure}
Fig.\ref{fig11} shows the changes in the total magnification of image as the function of the magnetic charge (left figure) for the different parameter values of the SIS parameter, and the SIS parameter  (right figure) for the different values of the magnetic charge where corresponding fixed parameters are $b=3$, $x_{0}=0.055$ and $R_{s}=2$. From Fig.\ref{fig11}, we can see that when we increase the SIS medium, the total magnification decreases gradually.
Because the plasma density decreases with the radius $(dN/dr<0)$, $\hat\alpha_{\text{sis3}}$ is negative which is opposite to the gravitational deflection (see Refs.\cite {Bisnovatyi-Kogan:2010flt} and  \cite{Bisnovatyi-Kogan:2015dxa}). If $\hat\alpha_{\text{sis3}}$ is  positive, the  total magnification of image as the function of $\omega_{c}^{2}/\omega^{2}$  has the opposite direction (see Refs.\cite{Babar:2021exh}). Fig.\ref{fig12} demonstrates the magnification ratio, i.e., the primary image $\mu_{+}^\text{pl}/\mu_{+}$ (lower curves) and the secondary image $\mu_{-}^\text{pl}/\mu_{-}$ (upper curves)  in the case we fix the  parameters as $q=-0.5$, $b=3$ and $R_{s}=2$. Because the effect of the  SIS  medium, the behavior of the magnification ratio is opposite to that of  the uniform plasma.

\subsection{Non-Singular isothermal sphere}
In this subsection, we focus on the total magnification and  the magnification ratio of image brightness for the  ESTGB  black hole surrounded by the NSIS medium. The Einstein angle $\theta_{0}^\text{pl}$ can be written as
\begin{equation}\label{Q52}
\begin{aligned}
(\theta^{pl}_{0})_\text{nsis}&=\theta_{0}\bigg\{\frac{1}{2}\bigg((2+\frac{8q^{3}}{3b^{2}R_{s}})+\big(\frac{R_{s}}{b\pi r_{c}^{2}}-\frac{b}{2(b^{2}+r_{c}^{2})^\frac{3}{2}}+\frac{b R_{s} \arctan{(\frac{r_{c}}{\sqrt{b^{2}+r_{c}^{2}}})}}{\pi r^{3}_{c}\sqrt{b^{2}+r_{c}^{2}}}\big)\\
&\times \frac{\omega_{c}^{2}R_{s} b}{\omega^{2}} +\big(-\frac{1}{b^{2} r_{c}^{4}} +\frac{2}{3 b^{4} r_{c}^{2}}+\frac{\arctan{(\frac{r_{c}}{\sqrt{b^{2}+r_{c}^{2}}})}}{r_{c}^{5}\sqrt{b^{2}+r_{c}^{2}}}\big)\frac{3 q^{3} b^{2} R_{s} \omega_{c}^{2}}{\omega^{2} \pi}\bigg)\bigg\}^{\frac{1}{2}}.
\end{aligned}
\end{equation}

The dimensionless parameter $x$ has the form
\begin{equation}\label{Q53}
\begin{aligned}
x_\text{nsis}&=\frac{\beta}{(\theta^{pl}_{0})_\text{nsis}}\\
& =x_{0}\bigg\{\frac{1}{2}\bigg((2+\frac{8q^{3}}{3b^{2}R_{s}})+\big(\frac{R_{s}}{b\pi r_{c}^{2}}-  \frac{b}{2(b^{2}+r_{c}^{2})^\frac{3}{2}}+\frac{b R_{s} \arctan{(\frac{r_{c}}{\sqrt{b^{2}+r_{c}^{2}}})}}{\pi r^{3}_{c}\sqrt{b^{2}+r_{c}^{2}}}\big)  \\
&\times\frac{\omega_{c}^{2}R_{s} b}{\omega^{2}}+\big(-\frac{1}{b^{2} r_{c}^{4}} +\frac{2}{3 b^{4} r_{c}^{2}}+\frac{\arctan{(\frac{r_{c}}{\sqrt{b^{2}+r_{c}^{2}}})}}{r_{c}^{5}\sqrt{b^{2}+r_{c}^{2}}}\big)\frac{3 q^{3} b^{2} R_{s} \omega_{c}^{2}}{\omega^{2} \pi}\bigg)\bigg\}^{-\frac{1}{2}},
\end{aligned}
\end{equation}
where $x_{0}=\beta/\theta_{0}$.

\begin{figure}[htbh]
	\centering
	{\includegraphics[width=0.46\textwidth,height=0.32\textwidth]{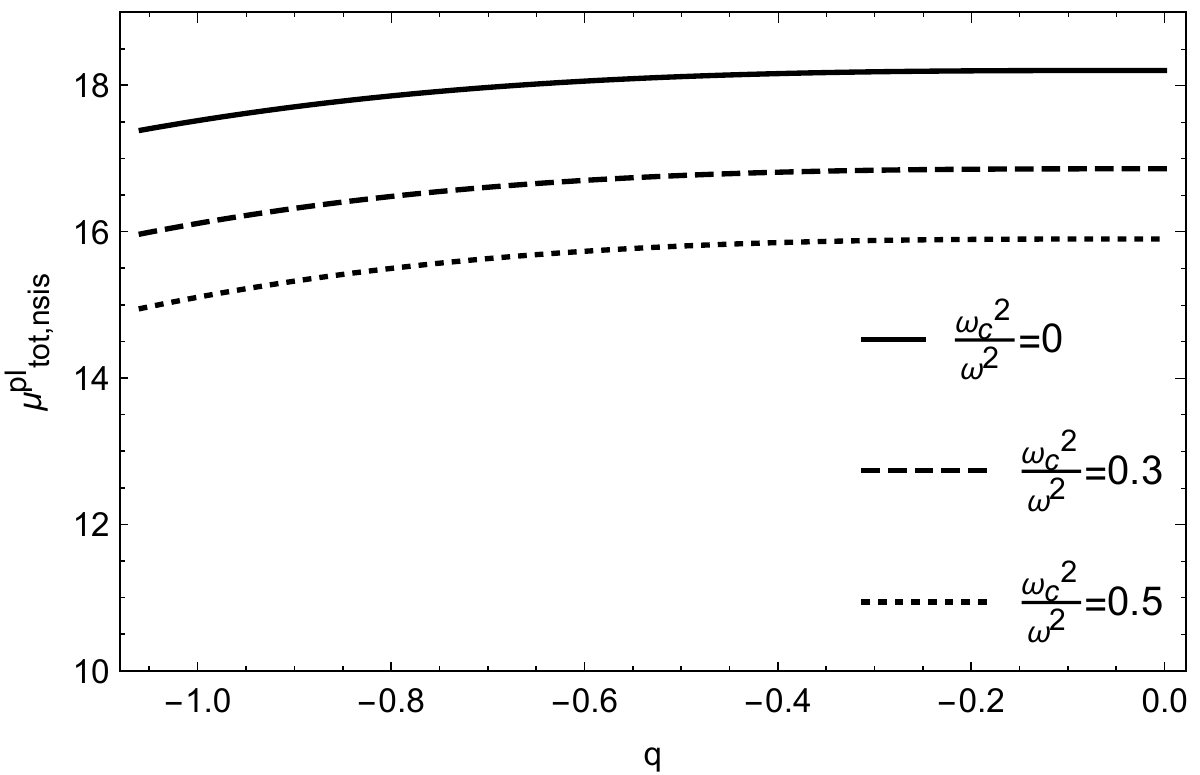}}
	{\includegraphics[width=0.46\textwidth,height=0.32\textwidth]{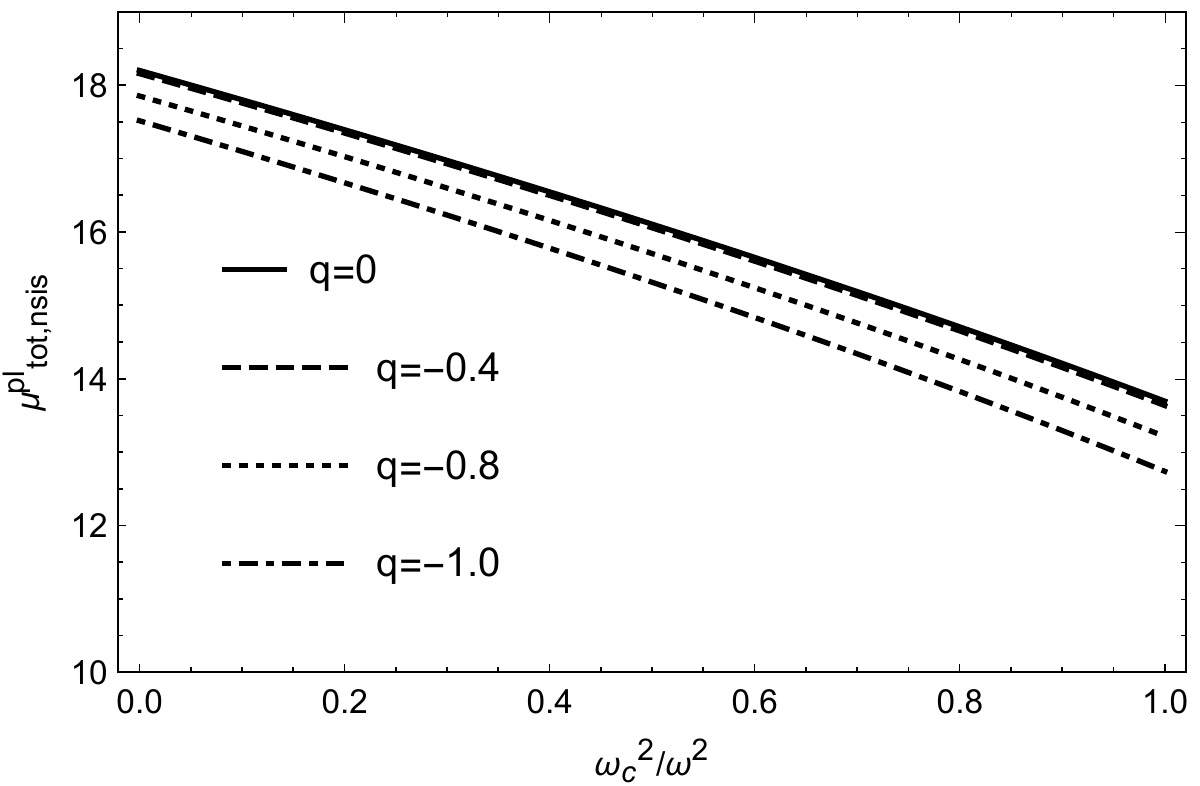}}
	\caption{The  total magnification of the image  brightness $\mu_{tot}$ as the function of  the
		magnetic charge (left panel),  NSIS parameter (right panel)  with  fixed $R_{s}=2$, $b=3$, $x_{0}=0.055$ and $r_{c}=3$.}
	\label{fig13}
\end{figure}
\begin{figure}[htbh]
	\centering
	{\includegraphics[width=0.48\textwidth,height=0.34\textwidth]{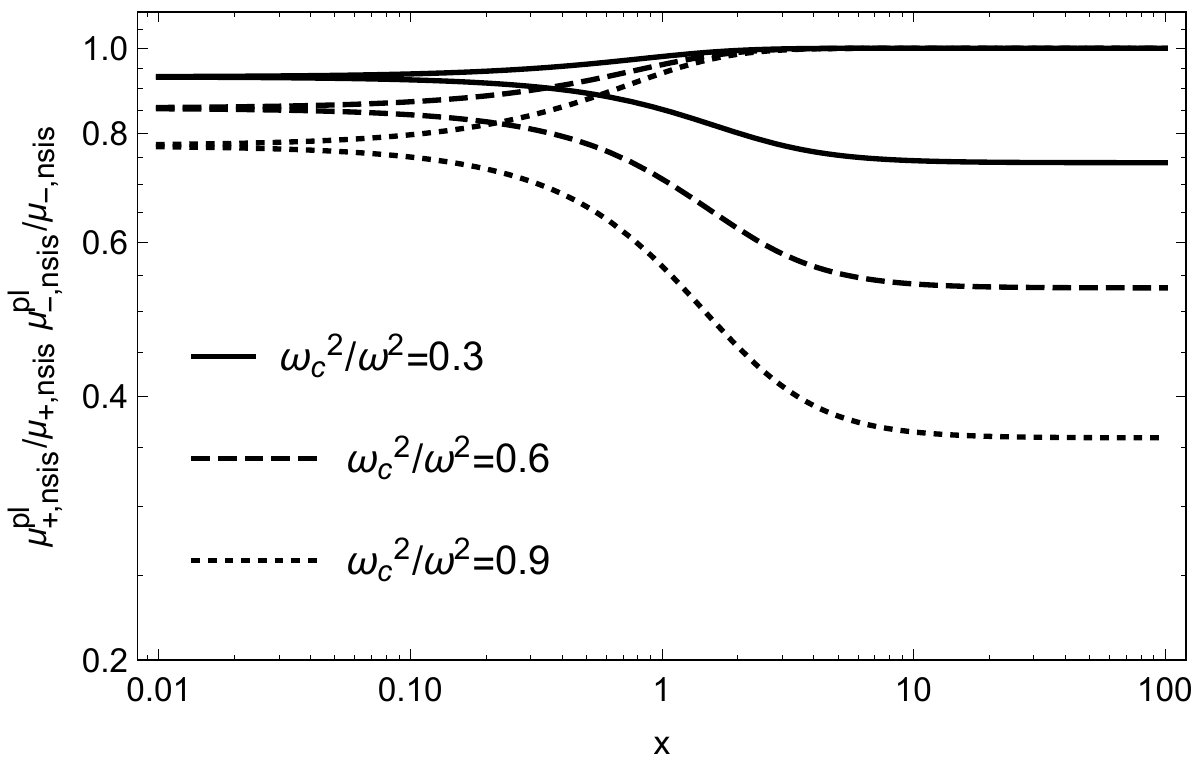}}
	\caption{ The ratio $\mu_{+}^\text{pl}/\mu_{+}$ (lower curves) and $\mu_{-}^\text{pl}/\mu_{-}$ (upper curves) of the magnification as the function of $x$  for the different values of the  NSIS parameter  distribution. We fix the  parameters as $q=-0.5$, $b=3$, $R_{s}=2$ and $r_{c}=3$.}
	\label{fig14}
\end{figure}

In Fig.\ref{fig13}, we show the graph of the total magnification for the case that the black hole is surrounded by the NSIS medium. By analyzing the behavior shown in  Fig.\ref{fig13}, one can see that the change is similar to the case of the  singular isothermal sphere. The presence of a NSIS reduces the total amplification in comparison with vacuum circumstance, i.e., $\omega_{c}^{2}/\omega^{2}=0$. This  is because  $\hat\alpha_{\text{nsis3}}$ is  negative. We also plot the changes of the magnification ratio of the primary and secondary images with  fixed $R_{s}=2$, $b=3$, $x_{0}=0.055$ and $r_{c}=3$ in Fig.\ref{fig14}. It is observed that  $\mu_{-}^\text{pl}/\mu_{-}$ (upper curves) tends to unity as larger $x$. And the ratio $\mu_{+}^\text{pl}/\mu_{+}$ (lower curves) is less than 1.
\begin{figure}[htbh]
	\centering
	{\includegraphics[width=0.48\textwidth,height=0.335\textwidth]{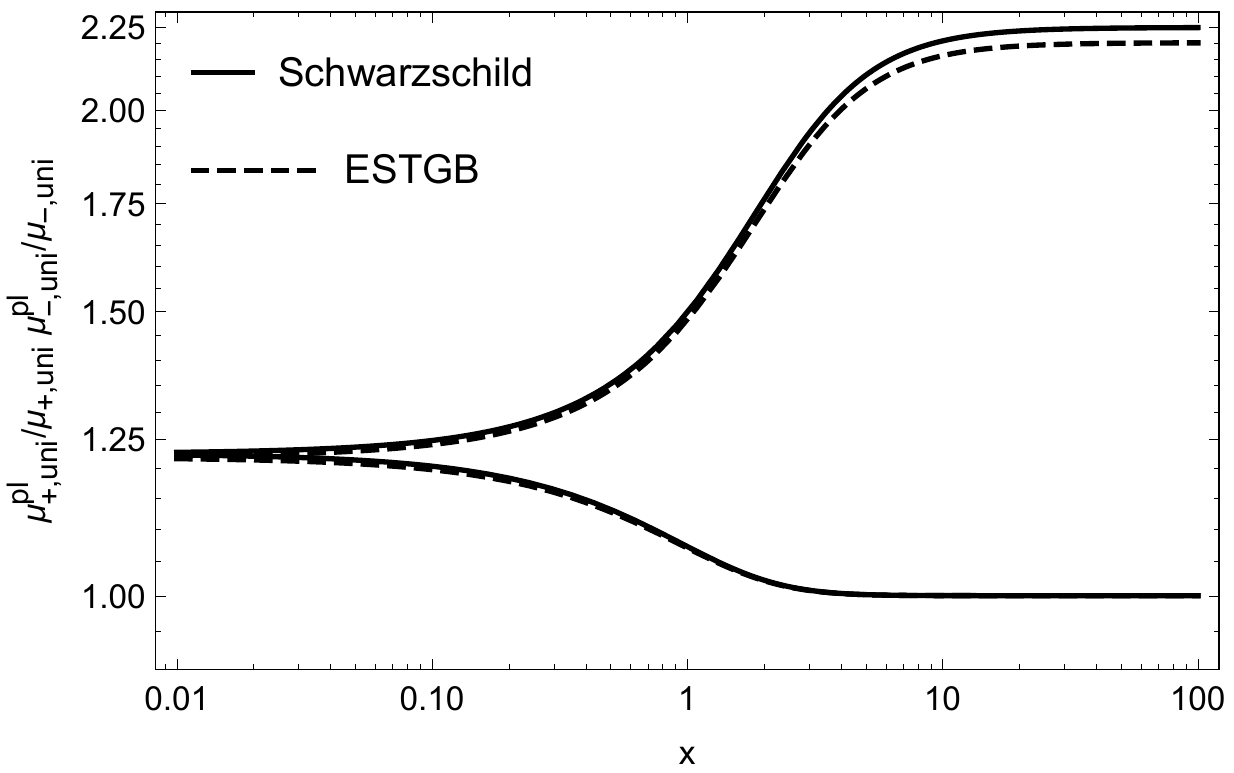}}
	{\includegraphics[width=0.48\textwidth,height=0.34\textwidth]{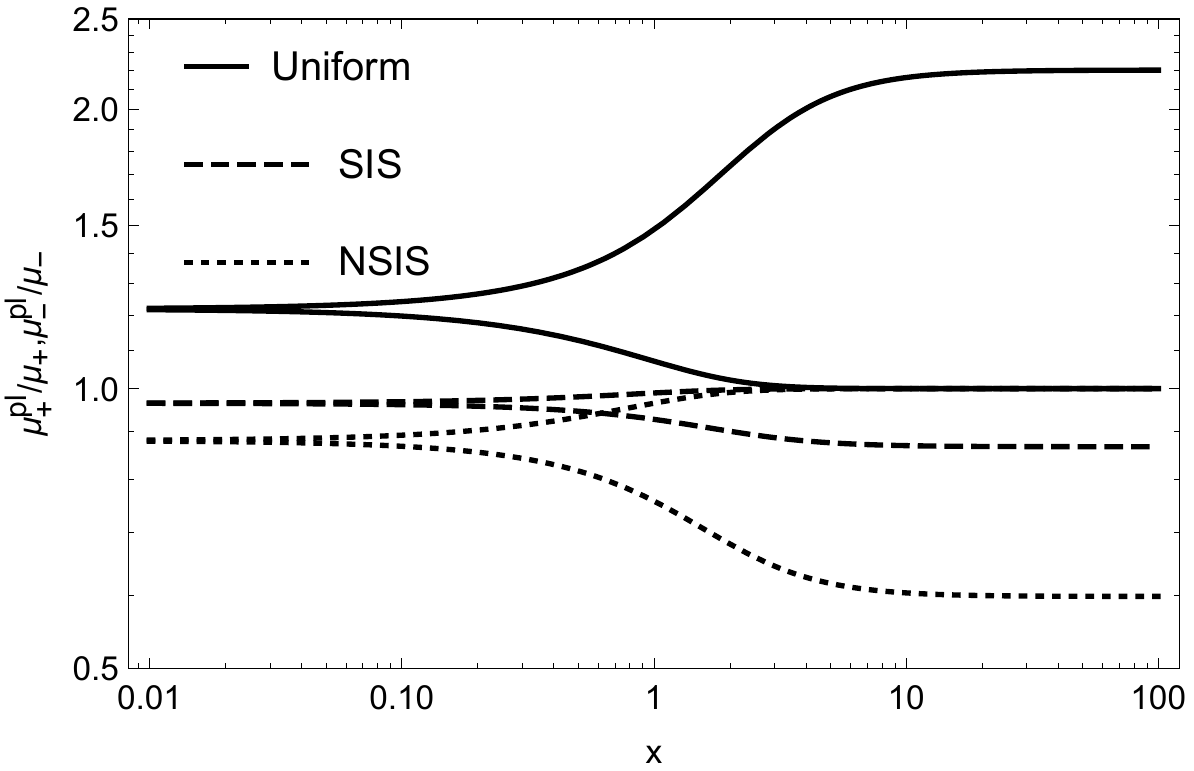}}
	\caption{The magnification ratio of image brightness of the Schwarzschild black hole and ESTGB black hole in the uniform plasma  (left figure); the ratio $\mu_{+}^\text{pl}/\mu_{+}$ (lower curves) and $\mu_{-}^\text{pl}/\mu_{-}$ (upper curves) of the magnification as the function of $x$  for uniform, SIS and NSIS  medium (right figure). We fix the  parameters as $q=-0.5$, $b=3$, $R_{s}=2$, $\omega_{0}^{2}/\omega^{2}=\omega_{c}^{2}/\omega^{2}=0.5$ and $r_{c}=3$.}
	\label{fig15}
\end{figure}

We compare the magnification ratio of image brightness of the Schwarzschild black hole and ESTGB black hole in the uniform plasma in Fig.\ref{fig15}.  We see that at large $x$ the  ratio of the magnification $\mu_{+}^\text{pl}/\mu_{+}$ tends to unity for the  Schwarzschild black hole and ESTGB black hole; the  ratio of the magnification $\mu_{-}^\text{pl}/\mu_{-}$ of the Schwarzschild black hole tends to a constant, 2.25. This is consistent with the results of Bisnovatyi-Kogan et al.\cite{Bisnovatyi-Kogan:2010flt}. In addition, the magnetic charge has slight influence on the magnification ratio of the image.

To compare the effects of the different plasma  models on  magnification ratio of image brightness, in Fig.\ref{fig15} we plot the magnification ratio of  the three plasma distributions, i.e., uniform, SIS and NSIS, with the same parameters  $q=-0.5$, $b=3$, $R_{s}=2$, $\omega_{0}^{2}/\omega^{2}=\omega_{c}^{2}/\omega^{2}=0.5$ and $r_{c}=3$.
We can obtain from Fig.\ref{fig15} that as a consequence of the non-uniform plasma distribution around the black hole, the  magnification ratio of the non-uniform plasma is less than that of uniform plasma. This means that only when there is uniform plasma around the black hole, the observer in the distance will perceive a considerable magnification.

\section{Conclusion and discussion}\label{4}
In the work, we discussed the weak gravitational lensing properties of a 4D  ESTGB  black hole immersed in  different plasma distribution models. We studied in detail the effect of the different plasma distribution models, i.e., uniform, SIS and NSIS medium, and the magnetic charge on the deflection of light. We found that the deflection angle increases slightly with the decrease of the absolute values of the magnetic charge. That is,  the black hole has the maximum deflection angle when it returns to the  Schwarzschild black hole. We showed that the presence of uniform plasma leads to an increase in the deflection angle. However, due to the fact that  $\hat\alpha_{\text{sis3}}$ $(\hat\alpha_{\text{\text{nsis3}}})$  caused by the plasma inhomogeneity  is less than zero , the deflection angle of the non-uniform plasma medium slightly diminishes with the increase of the plasma parameter. Moreover, compared with the SIS model, we found that the deflection angle is more sensitive to  parameters $b$ and  $\omega_{c}^{2}/\omega^{2}$ in the NSIS  model. We investigated the total magnification of  image due to the weak gravitational lensing effect around a plasma-surrounded black hole. We observed that the change of the total magnification is similar to that of the deflection angle. In other words, for the uniform plasma model, the magnification of image increases, while for SIS or NSIS model, the magnification of image decreases. This result is also indicated by the magnification ratio of the image source.
Finally, according to the influence of three plasma models on the deflection angle and the magnification of image, we can qualitatively understand the uniform plasma as a concave lens, while the SIS and NSIS plasma models as a convex lens in the context of the refractive index $n<1$.

%=========================================================================
\section*{Acknowledgments}
This work was supported partly by the National Natural Science Foundation of China (Grant No. 12065012), Yunnan High-level Talent Training Support Plan Young \& Elite Talents Project (Grant No. YNWR-QNBJ-2018-360) and the Fund for Reserve Talents of Young and Middle-aged Academic and Technical Leaders of Yunnan Province (Grant No. 2018HB006).

%\section{Reference}

\end{document}